\documentclass[aps,prb,twocolumn,showpacs,unsortedaddress,amsmath,amssymb,10pt]{revtex4-1}
\usepackage[utf8x]{inputenc}
\usepackage{amsmath}
\usepackage{amsfonts}
\usepackage{bbm}               
\usepackage[hyperfootnotes=false,pdfpagelabels,pagebackref,unicode]{hyperref}
\usepackage[caption=false]{subfig}
\usepackage{relsize}
\usepackage{placeins}
\usepackage{pstool}
\usepackage{bbold}
\usepackage{ulem}
\EndPreamble
\newcommand{\ket}[1]{\big|#1\big>}

\newcommand{\I}{ \mathrm{i} }
\newcommand{\E}{ \mathrm{e} }
\newcommand{\D}{ \mathrm{d} }
\newcommand{\thavg}[1]{\left\langle \right. \! #1 \! \left. \right\rangle} 
\makeatletter
\def\tagform@#1{\maketag@@@{\ignorespaces#1\unskip\@@italiccorr}}
\let\orgtheequation\theequation
\def\theequation{(\orgtheequation)}
\makeatother
\newsavebox{\mytmpbox}

\def\execB#1#2%
{\immediate
\write18{cp #1#2.eps}%
\write18{cp #1#2.tex}%
\psfragfig*{#2}
\endgroup
}

\newcommand{\img}[3]%
{{
\includegraphics[#1]{#2}
}}

\let\orgautoref\autoref
\renewcommand{\autoref}%
        {%
         \orgautoref}

\begin{document}
\bibliographystyle{utcaps}
\title{Single magnetic impurities in the Kane-Mele model}
\author{Florian Goth}
\author{David J. Luitz}
\author{Fakher F. Assaad}
\affiliation{Institut f\"ur Theoretische Physik und Astrophysik,\\
Universit\"at W\"urzburg, Am Hubland, D-97074 W\"urzburg, Germany}
\date{\today}

\begin{abstract}
The realization of the spin-Hall effect in quantum wells has led to
a plethora of studies regarding the properties of the edge states of a
two-dimensional topological insulator. These edge states constitute a class of one-dimensional liquids,
called the helical liquid, where an electron's spin quantization axis
is tied to its momentum. In contrast to one
dimensional conductors, magnetic impurities --- below the Kondo temperature --- cannot block transport and one expects the current to circumvent the impurity. To study this phenomenon, we consider the single impurity Anderson model embedded into an edge of a Kane-Mele
ribbon with up to $512\times80$ sites and use the numerically exact continuous time quantum Monte Carlo method (CTQMC) to
study the Kondo effect.
We present results on the temperature dependence of the spectral properties of the impurity and the bulk system
that show the behaviour of the system in the various regimes of the Anderson model. 
A view complementary to the single particle spectral functions can be obtained 
using the spatial behaviour of the spin spin correlation
functions. Here we show the characteristic, algebraic decay in the edge channel near the impurity.
\end{abstract}

\pacs{71.10.Fd, 05.10.Ln, 05.70.Ln}
\maketitle

\section{Introduction}
Numerical studies of variants of the Kane-Mele model\cite{PhysRevLett.95.226801} have recently been pursued
with increasing interest\cite{Hohenadler2012, Hohenadler2012_2, Thomas_Spin_Liquid, 2011PhRvL.106j0403H,PhysRevB.85.205102} since it can be used 
as a theoretical framework to study correlation effects in quantum spin-Hall insulators,
or two-dimensional (2D) topological insulators (TIs) \cite{2012arXiv1211.1774H}.
A characteristic feature is the formation of metallic edge states at the boundary of the system, which are robust to external perturbations
provided that time reversal symmetry is not broken. \cite{PhysRevB.73.045322}
They form a helical liquid, such that the electrons' spin is tied to its direction of motion \cite{PhysRevLett.96.106401}.

A particularly interesting perturbation of the helical edge state is the introduction of 
magnetic impurities interacting with the edge --- a problem usually modeled by an $S = \frac{1}{2}$ local spin that is coupled to the helical liquid. 
Due to the one-dimensional(1D) nature of the edge this problem has been studied extensively with bosonization techniques with variable Luttinger liquid parameter  accounting for correlation effects in the  helical edge state \cite{2009PhRvL.102y6803M, 2012arXiv1207.7081P, Maciejko2012, PhysRevB.81.041305}.
The Kondo effect in three dimensional TIs has been studied in Refs.~\cite{2010PhRvB..81w5411F, PhysRevB.81.035104, PhysRevB.81.233405, 2012arXiv1211.0034M}.
Returning to the 2D case and in the weak coupling regime with respect to electronic correlations on the edge, the formation of the Kondo singlet will effectively remove sites, thereby redefining the topology of the slab and the flow of the edge state.
The aim of this paper is to study the temperature dependence of this effect by computing, among other quantities, the site dependent density of states.
To do so,  we will set out to model the magnetic impurity using the single impurity Anderson model\cite{PhysRev.124.41}
which accounts for a single localized energy level that hybridizes with the states of the TI and has an
on-site Coulomb repulsion, while sites in the bulk are assumed to be noninteracting.
For non-vanishing Hubbard-interaction this model enables us to trace the progression from the high-temperature regime over the development of the local moment towards the formation of the Kondo singlet.
We achieve this with the numerically exact interaction expansion continuous time Quantum Monte Carlo
(CT-INT) algorithm introduced by Rubtsov et al.\cite{Rubtsov2005,2010arXiv1012.4474G,AssaadLang2007} which is particularly suitable
for the study of impurity problems, since noninteracting bath sites (the TI) do not count towards
the computational complexity of the algorithm and can be integrated out.
To access the single particle properties of the bath we calculate the self-energy on the impurity and then calculate the bath Green's functions using Dyson's equation.
This allows us to exhibit the deflection of the edge state at the impurity by looking at the spectral signatures arising in the bulk spectral functions due to the emerging Kondo effect on the impurity.
Towards the end we progress from the one-particle spectral functions to a two-particle quantity, the spatially resolved, equal-time,
spin spin correlation function. Using this quantity, we can inquire the spatial extent of the Kondo screening cloud in the bulk and along the edge.

\section{The model}
The Kane-Mele model was first proposed as a candidate for a possible quantum spin-Hall effect in graphene and its Hamiltonian is given by
\begin{equation}
 H_{\text{KM}} = H_t + H_\lambda,
\end{equation}
with
\begin{equation*}
 \begin{split}
  H_t&=-t\sum \limits_{\vec{i} \sigma} a_{\vec{i} \sigma}^\dagger (b^{\phantom{\dagger}}_{\vec{i} \sigma} + b^{\phantom{\dagger}}_{\vec{i} + \vec{a_1} - \vec{a_2}, \sigma} + b^{\phantom{\dagger}}_{\vec{i} - \vec{a_2}, \sigma}) + h.c.\\
  H_\lambda &=\lambda \sum \limits_{\vec{i} \sigma} \sigma \left[i a_{\vec{i} \sigma}^\dagger \left( a^{\phantom{\dagger}}_{\vec{i} + \vec{a_1},\sigma} + a^{\phantom{\dagger}}_{\vec{i} - \vec{a_2},\sigma}  + a^{\phantom{\dagger}}_{\vec{i} + \vec{a_2} - \vec{a_1},\sigma} \right) \right.\\
             &\left. -i b_{\vec{i} \sigma}^\dagger \left( b^{\phantom{\dagger}}_{\vec{i} + \vec{a_1},\sigma} + b^{\phantom{\dagger}}_{\vec{i} - \vec{a_2},\sigma}  + b^{\phantom{\dagger}}_{\vec{i} + \vec{a_2} - \vec{a_1},\sigma}\right)\right] + h.c.
 \end{split}
\end{equation*}
 $a_{\vec{i}\sigma}$ and $b_{\vec{i}\sigma}$ denote fermionic operators acting on the respective sublattice and, to account for edge states, we consider this model on a slab geometry.
The boundary conditions are periodic in $r$-direction -- the corresponding number of sites is denoted by $N_x$ -- and open in $n$-direction with a length of $N_y$ sites.
$\lambda$ is the strength of the spin-orbit interaction and the hopping $t$ is set to $t = 1$ for everything that follows.
The effect of electron electron interactions in graphene was studied in Ref.\cite{PhysRevB.81.115416, 2011PhRvB..83s5432L,2012PhRvB..86g5458S}.
Although it turned out that in graphene the spin-orbit coupling is too small to observe the QSH state, the model can still be used as an effective Hamiltonian for this topological state of matter.
\begin{figure}
\def\svgwidth{\linewidth}
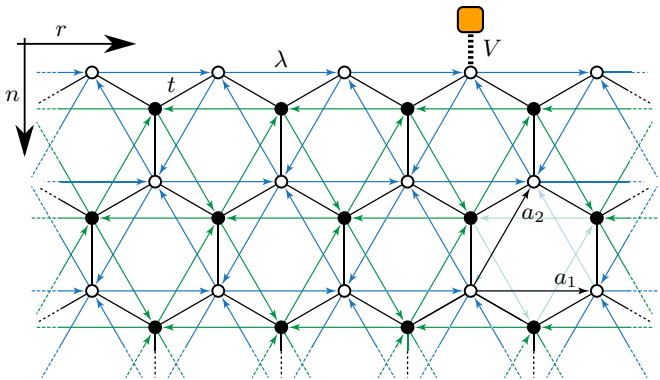
\caption{(Color online) The ribbon is periodic along the $r$ - direction and at site $r=0$,
$n=0$ the orange box denotes the impurity orbital which couples with a matrix element $V$. 
Sublattice $A$ is denoted by open circles and sublattice $B$ by filled circles. $a_1$ and $a_2$ denote the unit vectors of the honeycomb lattice.}
\end{figure}
This model is related to the spinless Haldane model that shows a quantum Hall effect but breaks time reversal invariance (TRI) \cite{Fiete2012845}.
The Kane-Mele model can be understood as two copies of the Haldane model while preserving time-reversal symmetry and exhibiting a quantum spin-Hall effect.
\begin{figure}
\includegraphics[width=\linewidth]{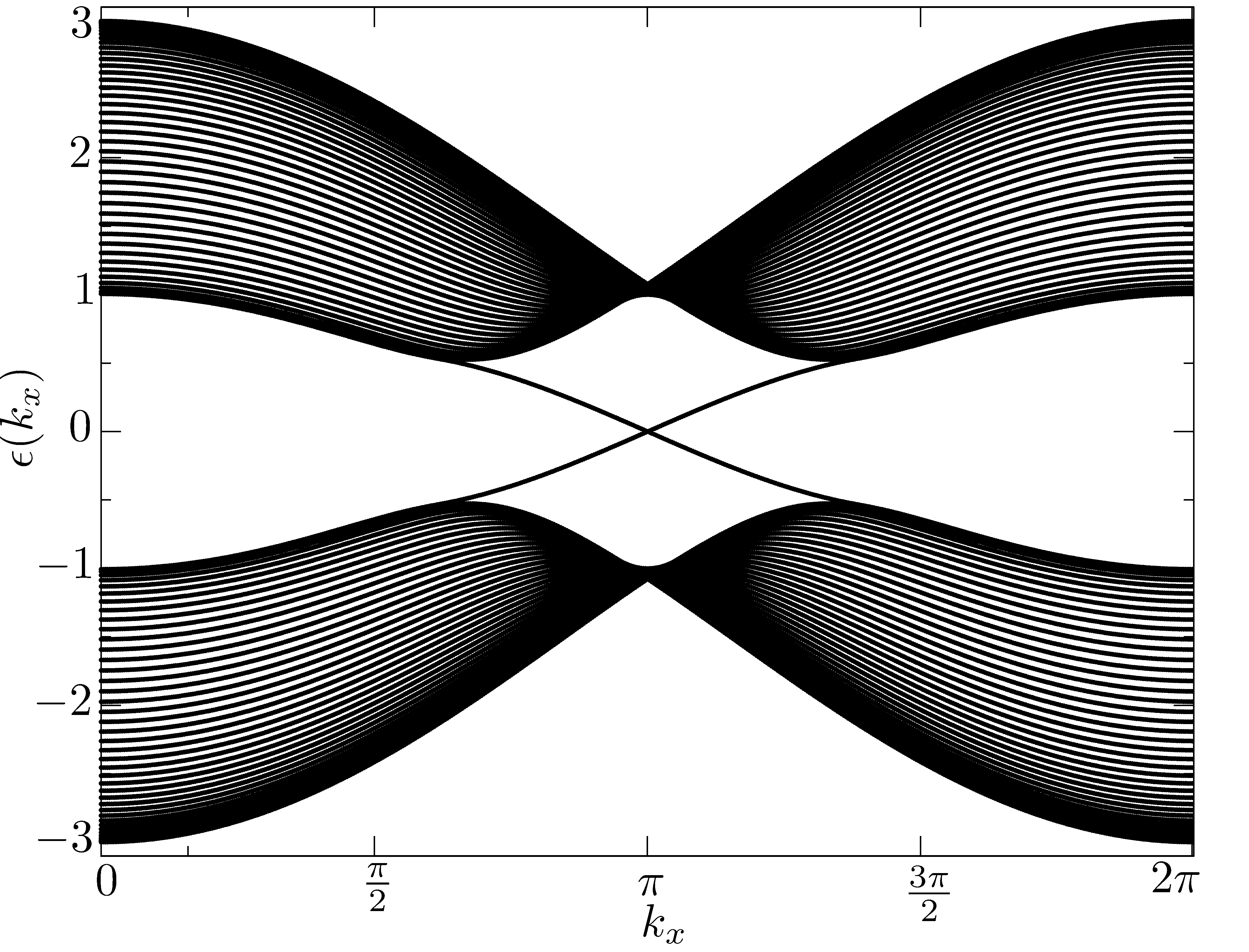}
\caption
{%
Spectrum of the Kane-Mele model. Here we have used $N_x=512$, $N_y=80$ and $\lambda = 0.1$, which constitute our ``canonical values'' in the following.
Visible are the different bands due to the "orbitals" in $n$-direction, as well as the famous edge states crossing at the Fermi energy.
}
\label{fig:KMspectra}
\end{figure}
Into this bath system we embed an impurity at an edge. The impurity's Hamiltonian $H_{\text{imp}}$ is given by
\begin{equation}
 H_ {\text{imp}}= H_0 + H_U
\end{equation}
with
\begin{equation*}
\begin{split}
H_0 &= \epsilon_d \sum \limits_\sigma d_\sigma^\dagger d^{\phantom{\dagger}}_\sigma + V \sum \limits_\sigma (a^\dagger_{\vec{0},\sigma}d^{\phantom{\dagger}}_\sigma + d^\dagger_\sigma a^{\phantom{\dagger}}_{\vec{0},\sigma}) \\
H_U &= U\left(n_\uparrow^d-\frac{1}{2}\right)\left(n_\downarrow^d-\frac{1}{2}\right).
\end{split}
\end{equation*}
Here, $\epsilon_d$ denotes the energy of the dot, $U$ the Hubbard interaction and $V$ the hybridization between the  first bath site and the impurity. 
 $d_\sigma$ denotes fermionic operators acting on the impurity. 
 We have chosen a symmetric representation of the Hubbard interaction that sets the chemical potential to zero for the half-filled case.
 We note that the impurity Hamiltonian obeys time reversal symmetry together with the bath.

\section{Summary of bath properties}
\begin{figure}
\hspace{-1em}
\def\svgwidth{1.1\linewidth}
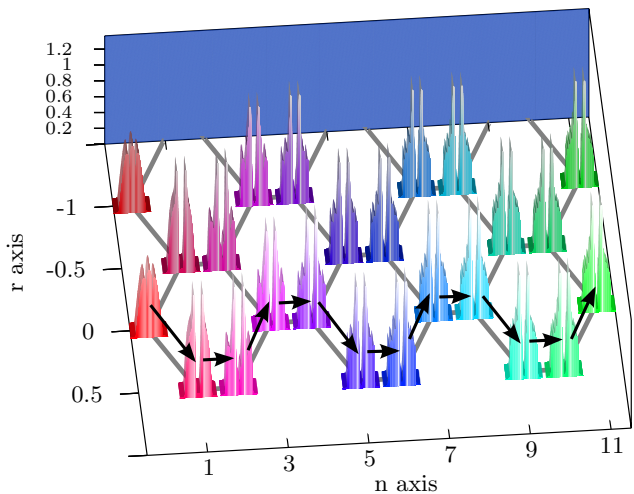
\caption{(Color online) This plot shows a part of the real-space lattice at $\lambda = 0.1$ where to every lattice point we have attached the spectral function $A^0_n(\omega)$.
For a given value of $r$, the black arrows denote the path that is taken through the lattice if $n$ is increased. Since the impurity is not yet added 
we have translation invariance along the $r$ direction. The bulge of the edge state in the $n=0$ spectral functions is clearly visible.}
\label{fig:bath_spectra}
\end{figure}

As already mentioned, the bath model $H_{\text{KM}}$ exhibits the so-called edge states which, as the name implies, are localized at the edges.
Since we attach the impurity to a site belonging to an edge we revisit some properties of the bath.
We refer to anything outside of the impurity as bath and everything in the bath that is not dominated by the edge state as bulk.
The edge states correspond to the states in the energy spectrum of \autoref{fig:KMspectra} that cross at the Fermi energy, $\epsilon(k_x) = 0$, and enable gapless electronic excitations at the edge.
These edge states constitute a helical liquid where the spin of an electron is coupled to the direction of propagation, hence 
an interaction flipping the spin reverses its momentum.
Since we will argue quite a bit with the help of the spectral functions we point out the general structure of $A_n(r,\omega)$ here.
It is
\begin{equation}
 A_n(r,\omega) = A^0_n(\omega) + B_n(r,\omega, V) + C_n(r, \omega, \Sigma(\omega))
 \label{eq:structurespectralfunction}
\end{equation}
with the impurity independent background $A^0_n(\omega)$, a term $B_n(r,\omega, V)$ that depends on the hybridization $V$ between lattice and impurity
and the contribution $C_n(r, \omega, \Sigma(\omega))$ due to the self-energy $\Sigma(\omega)$ of the impurity.
In \autoref{fig:bath_spectra} we show a site-resolved view onto the spectral functions $A^0_n(\omega)$ of the bath.
\begin{figure}
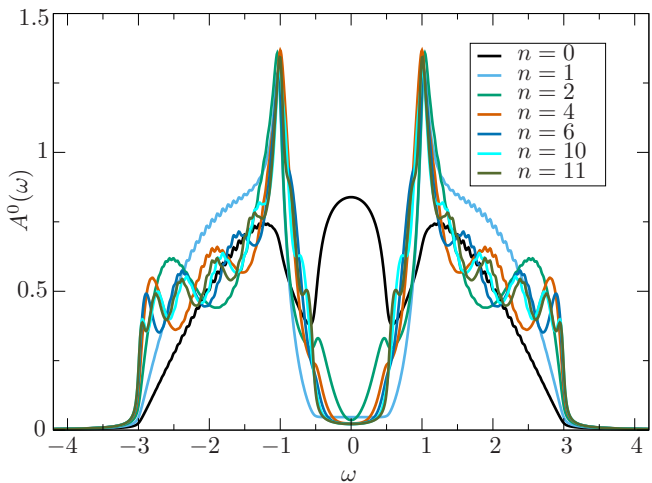

\img{width=\linewidth}{./free_spectral_functions}
{\psfrag{om}[Bl][Bl][1.9]{$\omega$}
\psfrag{Aom}[Bl][Bl][1.9]{$\hspace{1em}A^0(\omega)$}
\psfrag{0}[Bl][Bl][1.9]{$0$}
\psfrag{1}[Bl][Bl][1.9]{$1$}
\psfrag{2}[Bl][Bl][1.9]{$2$}
\psfrag{3}[Bl][Bl][1.9]{$3$}
\psfrag{4}[Bl][Bl][1.9]{$4$}
\psfrag{ r = 0, n = 0}[Bl][Bl][1.85]{$n=0$}
\psfrag{ r = 0, n = 1}[Bl][Bl][1.85]{$n=1$}
\psfrag{ r = 0, n = 2}[Bl][Bl][1.85]{$n=2$}
\psfrag{ r = 0, n = 4}[Bl][Bl][1.85]{$n=4$}
\psfrag{ r = 0, n = 6}[Bl][Bl][1.85]{$n=6$}
\psfrag{ r = 0, n = 10}[Bl][Bl][1.85]{$n=10$}
\psfrag{ r = 0, n = 11}[Bl][Bl][1.85]{$n=11$}
\psfrag{0.5}[Bl][Bl][1.9]{$\hspace{-2pt}0.5$}
\psfrag{1.5}[Bl][Bl][1.9]{$\hspace{-2pt}1.5$}
\psfrag{-1}[Bl][Bl][1.9]{$-1$}
\psfrag{-2}[Bl][Bl][1.9]{$-2$}
\psfrag{-3}[Bl][Bl][1.9]{$-3$}
\psfrag{-4}[Bl][Bl][1.9]{$\hspace{-0.6em}-4$}
\psfrag{-1}[Bl][Bl][1.9]{$\hspace{-0.2cm}-1$}
\psfrag{-2}[Bl][Bl][1.9]{$\hspace{-0.2cm}-2$}
\psfrag{-3}[Bl][Bl][1.9]{$\hspace{-0.2cm}-3$}}
\caption{(Color online) Spectral functions of \autoref{fig:bath_spectra} for comparison in a more traditional 2D plot.
Both plots are from a system with $N_x=512$ and $N_y = 80$ sites at $\lambda=0.1$. The fine wiggles in the spectral functions, as e.g. in $n=0$ or $n=1$, are artifacts of the finite system size. The edge state at $n = 0$ is clearly visible which almost immediately decays farther in the bulk.
Having used a finite $\eta = 4 \Delta \omega \approx 0.03$ we have some broadening which prevents the gap of the spectral function from vanishing in the bulk. $\Delta \omega$ is the resolution of the energy $\omega$.
}
\label{fig:bath_cut}
\end{figure}
The bulge that is visible in the outermost ($n=0$) spectral functions is the edge state,
which has its spectral weight centered around $\omega = 0$.
Since we consider the system without an impurity we have translation invariance along the $r$ - direction.
For comparison we show in \autoref{fig:bath_cut} a cut along $r = 0$ of the same spectral functions.
Further into the bulk the gap of the insulator appears. Also we see the odd-even pattern close to the edge.
The $n = 1$ spectral function shows a gap, whereas the $n=2$ function shows some remains of the 
exponentially decaying edge state.

\section{An uncorrelated impurity}
\begin{figure*}
\begin{tabular}{cc}
\subfloat[\strut]{%
 \def\svgwidth{0.55\linewidth}
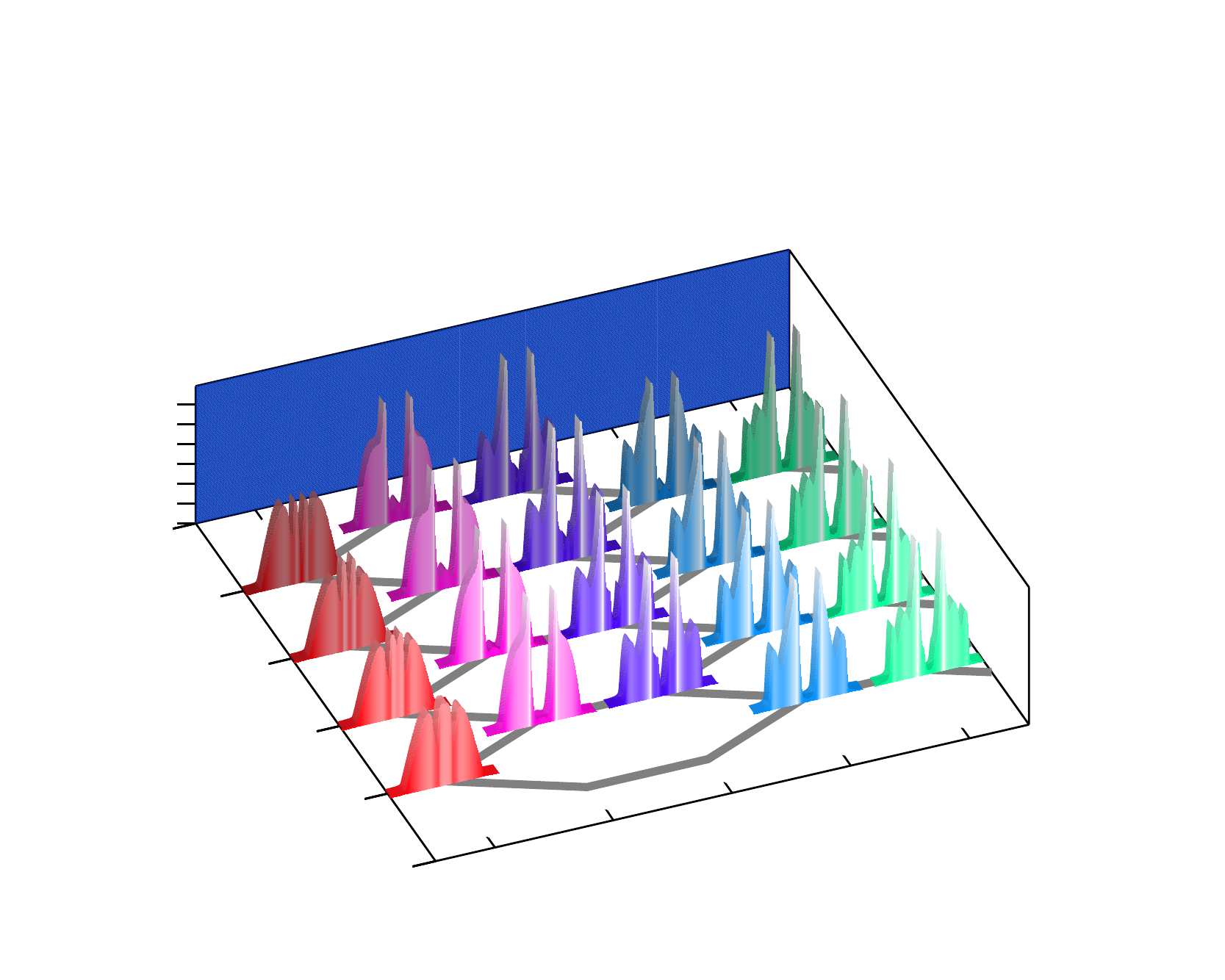
\label{fig:fullspectra_imp_no_correlation}
}&
\hspace{-0.04\linewidth}\subfloat[\strut]{%
\img{width=0.44\linewidth}{./full_spectralfunctions_no_corr}
{\psfrag{om}[Bl][Bl][1.0]{$\omega$}
\psfrag{Aom}[Bl][Bl][1.0]{$A(\omega)$}
\psfrag{0}[Bl][Bl][1.0]{$0$}
\psfrag{1}[Bl][Bl][1.0]{$1$}
\psfrag{2}[Bl][Bl][1.0]{$2$}
\psfrag{3}[Bl][Bl][1.0]{$3$}
\psfrag{4}[Bl][Bl][1.0]{$4$}
\psfrag{-1}[Bl][Bl][1.0]{$\hspace{-0.5em}-1$}
\psfrag{-2}[Bl][Bl][1.0]{$\hspace{-0.5em}-2$}
\psfrag{-3}[Bl][Bl][1.0]{$\hspace{-0.5em}-3$}
\psfrag{-4}[Bl][Bl][1.0]{$\hspace{-0.5em}-4$}
\psfrag{0.5}[Bl][Bl][1.0]{\hspace{-0.4em}$0.5$}
\psfrag{1.5}[Bl][Bl][1.0]{\hspace{-0.4em}$1.5$}
\psfrag{nnn = 0}[Bl][Bl][1.0]{$n=0$}
\psfrag{n = 1}[Bl][Bl][1.0]{$n=1$}
\psfrag{n = 2}[Bl][Bl][1.0]{$n=2$}
\psfrag{n = 3}[Bl][Bl][1.0]{$n=3$}
\psfrag{n = 4}[Bl][Bl][1.0]{$n=4$}
\psfrag{n = 5}[Bl][Bl][1.0]{$n=5$}}
\label{fig:fullspectra_imp_no_correlation_cut}
}\\
 \subfloat[\strut
 ]{%
 \def\svgwidth{0.55\linewidth}
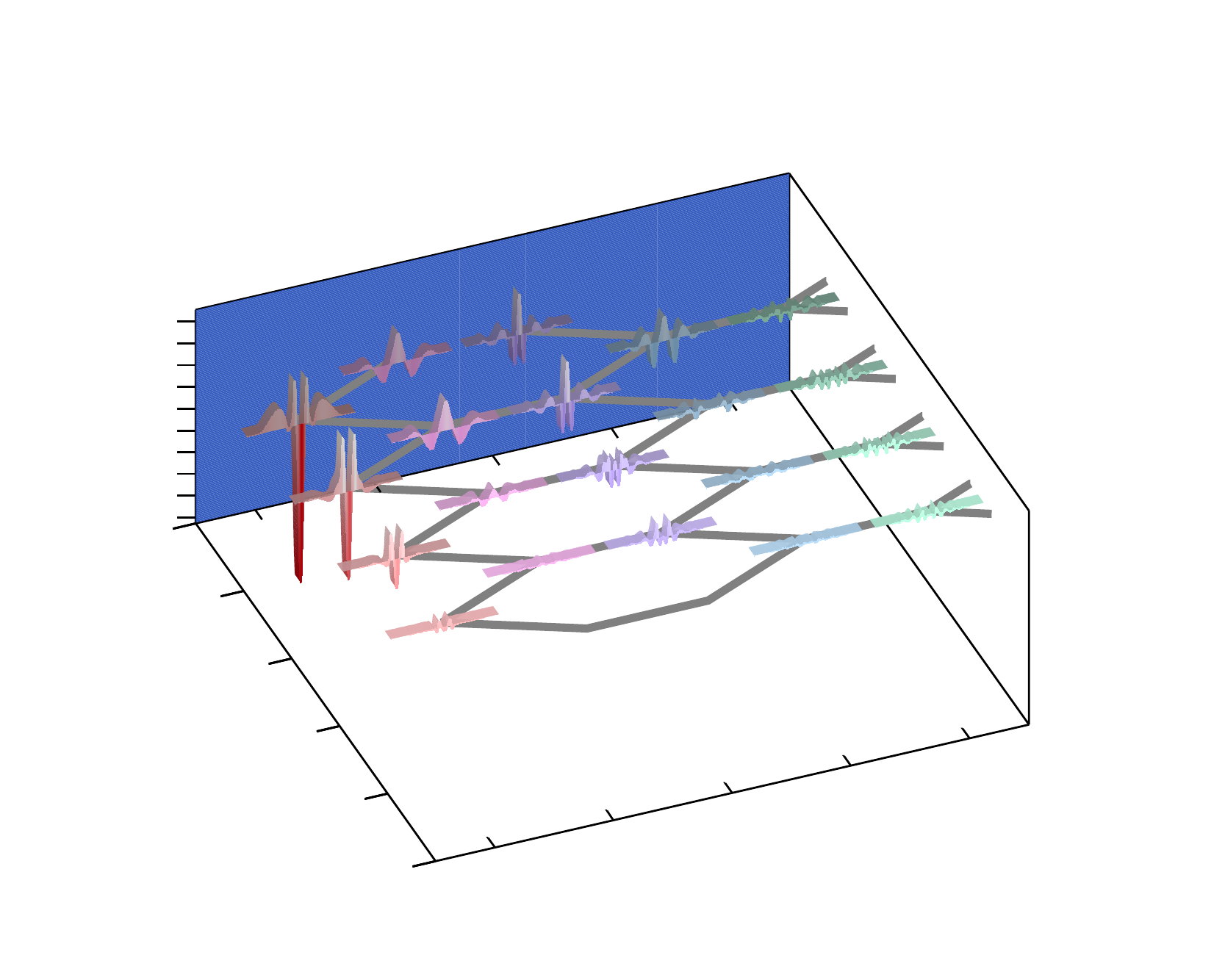
 \label{fig:hyb_effect_overview}}&
\hspace{-0.04\linewidth}\subfloat[\strut]{%
\img{width=0.44\linewidth}{./deltaA_no_corr}{
\psfrag{dAom}[Bl][Bl][1.0]{$B_n(0,\omega, V)$}
\psfrag{om}[Bl][Bl][1.0]{$\omega$}
\psfrag{0}[Bl][Bl][1.0]{$0$}
\psfrag{1}[Bl][Bl][1.0]{$1$}
\psfrag{2}[Bl][Bl][1.0]{$2$}
\psfrag{3}[Bl][Bl][1.0]{$3$}
\psfrag{4}[Bl][Bl][1.0]{$4$}
\psfrag{-1}[Bl][Bl][1.0]{$\hspace{-0.5em}-1$}
\psfrag{-2}[Bl][Bl][1.0]{$\hspace{-0.5em}-2$}
\psfrag{-3}[Bl][Bl][1.0]{$\hspace{-0.5em}-3$}
\psfrag{-4}[Bl][Bl][1.0]{$\hspace{-0.5em}-4$}
\psfrag{nnn = 0}[Bl][Bl][1.0]{$n=0$}
\psfrag{n = 1}[Bl][Bl][1.0]{$n=1$}
\psfrag{n = 2}[Bl][Bl][1.0]{$n=2$}
\psfrag{n = 3}[Bl][Bl][1.0]{$n=3$}
\psfrag{n = 4}[Bl][Bl][1.0]{$n=4$}
\psfrag{n = 5}[Bl][Bl][1.0]{$n=5$}
\psfrag{0.2}[Bl][Bl][1.0]{\hspace{-0.3em}$0.2$}
\psfrag{-0.2}[Bl][Bl][1.0]{\hspace{-1em}$-0.2$}
\psfrag{-0.4}[Bl][Bl][1.0]{\hspace{-1em}$-0.4$}
\psfrag{-0.6}[Bl][Bl][1.0]{\hspace{-1em}$-0.6$}
\psfrag{-0.8}[Bl][Bl][1.0]{\hspace{-1em}$-0.8$}
}\label{fig:hyb_effect_cut}}
\end{tabular}
\caption{(Color online) (a) shows the complete spectral functions now with a non-interacting impurity located at $r=0$ and $n = 0$.
The original edge state is deformed into the bulk around the impurity. The missing weight in the $n=0$ spectral functions 
shows up in the $n=1, 2$ spectral functions.
(b) is a cut of $A_n(r, \omega)$ along $r = 0$ that shows the rearrangement of spectral weight from the edge into the bulk.
(c) shows $B_n(r,\omega, V)$, the effect on the spectral functions attributable to the hybridization $V$ of the impurity
with the edge state.
 We see large negative contributions to the edge state at the impurity and positive contributions farther into the bulk.
 $B_n(r = 0,\omega, V)$ is shown in (d).
 Obvious is the strong reduction at the site below the impurity $(r=0, n=0)$ which is shifted to the sites around the impurity.
}
\end{figure*}

Adding to the bath system the uncorrelated impurity given by $H_0$ at site $r = 0$ and $n = 0$ the spectral
properties of the system change around the impurity since now the $B$ - term in \autoref{eq:structurespectralfunction} contributes to the 
spectral functions due to the hybridization $V$.
From \autoref{fig:fullspectra_imp_no_correlation} and \autoref{fig:fullspectra_imp_no_correlation_cut} it is visible that right at the impurity the
spectral weight of the edge state is drastically reduced. 
In contrast to the changes due to correlations in \autoref{fig:change_nocorr} we see that locally this potential poses quite 
a strong perturbation to $A(\omega)$. Since time reversal symmetry is present, single particle backward scattering is prohibited as this would amount to 
flipping the orientation of the spin. As a consequence, the edge state has to circumvent the potential impurity by deflecting into the bulk.
Thus the missing spectral weight reappears at sites further into the bulk.
\autoref{fig:hyb_effect_overview} and \autoref{fig:hyb_effect_cut} show the change in the spectral function, $B_n(r,\omega, V)$, due to the hybridization.
The edge state is deformed around the impurity since it acts as a pure potential scatterer for the edge state.
This is consistent with the spectral function of the impurity which is just a lorentzian around $\omega = 0$
similar to the ``correlated'' spectral function at $\beta= 0.1$ in \autoref{fig:dot_spectra_U2}.
Although the edge state is protected by symmetry against potential scattering, the effect of the impurity is that 
it acts as a trap for electrons from the bulk which can then in turn interact with the electrons of the edge state \cite{2009PhRvL.102y6803M}.
An interpretation of the resulting new path of the edge channel is that the site to which the impurity is
connected is effectively removed from the lattice.
In that sense the deformation of the edge state can be understood as a rerouting along the changed edge of the system.
A similar deflection of the edge state around centers of potential scattering has also been reported in 3D TI's. \cite{2012arXiv1203.2628S}

\section{A correlated impurity}
\begin{figure}
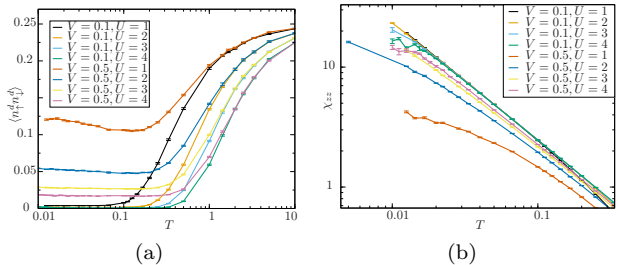

 \subfloat[\relax]
 {%
 \img{width=0.45\linewidth}{./docc_comparison}
{\psfrag{T}[Bl][Bl][1.0]{$T$}
\psfrag{nupndown}[Bl][Bl][1.0]{$\langle n^d_\uparrow n^d_\downarrow \rangle$}
\psfrag{0}[Bl][Bl][1.0]{$\hspace{-0.5em}0$}
\psfrag{10}[Bl][Bl][1.0]{$\hspace{-0.6em}10$}
\psfrag{1}[Bl][Bl][1.0]{$1$}
\psfrag{0.1}[Bl][Bl][1.0]{\hspace{-0.6em}$0.1$}
\psfrag{0.01}[Bl][Bl][1.0]{$0.01$}
\psfrag{0.05}[Bl][Bl][1.0]{\hspace{-0.6em}$0.05$}
\psfrag{0.15}[Bl][Bl][1.0]{}
\psfrag{0.2}[Bl][Bl][1.0]{$\hspace{-0.6em}0.2$}
\psfrag{0.25}[Bl][Bl][1.0]{$\hspace{-0.6em}0.25$}
\psfrag{VVVV = 0.1 , U = 1}[Bl][Bl][1.0]{$V=0.1, U=1$}
\psfrag{V=0.1, U=2}[Bl][Bl][0.98]{$V=0.1, U=2$}
\psfrag{V=0.1, U=3}[Bl][Bl][0.98]{$V=0.1, U=3$}
\psfrag{V=0.1, U=4}[Bl][Bl][0.98]{$V=0.1, U=4$}
\psfrag{V=0.5, U=1}[Bl][Bl][0.98]{$V=0.5, U=1$}
\psfrag{V=0.5, U=2}[Bl][Bl][0.98]{$V=0.5, U=2$}
\psfrag{V=0.5, U=3}[Bl][Bl][0.98]{$V=0.5, U=3$}
\psfrag{V=0.5, U=4}[Bl][Bl][0.98]{$V=0.5, U=4$}
}%
\label{fig:docc_comparison}
 }%
 \subfloat[\relax]
 {
\img{width=0.45\linewidth}{./zsusceptibility_new}
{\psfrag{T}[Bl][Bl][1.0]{$T$}
\psfrag{chi}[Bl][Bl][1.0]{$\chi_{zz}$}
\psfrag{0}[Bl][Bl][1.0]{$0$}
\psfrag{0.5}[Bl][Bl][1.0]{$0.5$}
\psfrag{1.5}[Bl][Bl][1.0]{$1.5$}
\psfrag{0.1}[Bl][Bl][1.0]{$0.1$}
\psfrag{0.01}[Bl][Bl][1.0]{$0.01$}
\psfrag{1}[Bl][Bl][1.0]{$\hspace{-0.8em}1$}
\psfrag{2}[Bl][Bl][1.0]{$2$}
\psfrag{3}[Bl][Bl][1.0]{$3$}
\psfrag{4}[Bl][Bl][1.0]{$4$}
\psfrag{5}[Bl][Bl][1.0]{$5$}
\psfrag{10}[Bl][Bl][1.0]{$\hspace{-0.8em}10$}
\psfrag{15}[Bl][Bl][1.0]{$15$}
\psfrag{20}[Bl][Bl][1.0]{$20$}
\psfrag{VVVV = 0.1, U = 1}[Bl][Bl][1.0]{$V=0.1, U=1$}
\psfrag{V = 0.1, U = 2}[Bl][Bl][1.0]{$V=0.1, U=2$}
\psfrag{V = 0.1, U = 3}[Bl][Bl][1.0]{$V=0.1, U=3$}
\psfrag{V = 0.1, U = 4}[Bl][Bl][1.0]{$V=0.1, U=4$}
\psfrag{V = 0.5, U = 1}[Bl][Bl][1.0]{$V=0.5, U=1$}
\psfrag{V = 0.5, U = 2}[Bl][Bl][1.0]{$V=0.5, U=2$}
\psfrag{V = 0.5, U = 3}[Bl][Bl][1.0]{$V=0.5, U=3$}
\psfrag{V = 0.5, U = 4}[Bl][Bl][1.0]{$V=0.5, U=4$}
}%
\label{fig:chi_zz}
}
\caption{(Color online) Overview of, (a), the double occupancy and (b), $\chi_{zz}$, as a function of temperature $T = \frac{1}{\beta}$.
The plots of the double occupancy highlight the different regimes. All functions start out at the uncorrelated value $\langle n_\uparrow n_\downarrow \rangle = 0.25$ and then fall towards some dip at intermediate temperatures. This dip roughly coincides with the local moment regime.
After that, the double occupancy increases slightly up to a saturation which is a signature of the Kondo regime.
}
\end{figure}

Now we add the Hubbard interaction $H_U$ to the impurity which leads us to consider the single impurity Anderson model (SIAM)
\begin{equation}
 H=H_{\text{imp}} + H_{\text{KM}}
\end{equation}
with the bath given by the Kane-Mele model $H_{\text{KM}}$.
Integrating out the bath electrons we obtain the action
\begin{equation}
\begin{split}
 S &= -\sum \limits_{\sigma} \int \limits_{0}^\beta d\tau \int \limits_{0}^\beta d\tau' d_\sigma^\dagger (\tau) G^{-1}_{d,d}(\tau - \tau') d^{\phantom{\dagger}}_\sigma(\tau')\\
 &+ U \int \limits_0^\beta d\tau \left(n_\uparrow^d(\tau) - \frac{1}{2}\right)\left(n_\downarrow^d(\tau) - \frac{1}{2}\right)
 \end{split}
\label{eq:dot_action}
\end{equation}
which is perfectly suitable for an implementation of a numerically exact CTQMC method that expands in the interaction strength $U$.
The computational effort is reduced if one uses the property that our model is time reversal invariant which leads to a spin-diagonal 
impurity Green's function $G_{d,d}$ as detailed in \autoref{sec:Gisdiagonal} for a general time reversal symmetric Hamiltonian. 
To at best study the temperature dependence of the above impurity problem, we will compute several quantities.  
The double occupancy, $\langle n^d_\uparrow n^d_\downarrow \rangle$ in \autoref{fig:docc_comparison} will allow us to track the
formation of the local moment and its screening. The same information is essentially contained in the local spin susceptibility 
\begin{equation*}
 \chi_{zz} = \int \limits_0^\beta d\tau \langle S^z(\tau) S^z \rangle
\end{equation*}
measured on the impurity and plotted in \autoref{fig:chi_zz}.
Due to TRI it is sufficient to consider only $\chi_{zz}$ since the other components are degenerate.

\begin{figure}
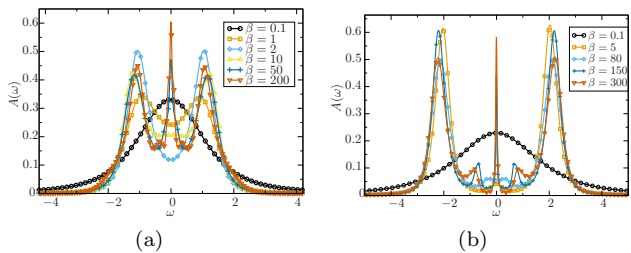

\hspace{-3em}
\subfloat[\strut]{
   \img{width=0.46\linewidth}{./dot_spectra_U2}
{\psfrag{om}[Bl][Bl][1.0]{$\omega$}
\psfrag{Aom}[Bl][Bl][1.0]{$A(\omega)$}
\psfrag{beta=0.1}[Bl][Bl][1]{$\beta = 0.1$}
\psfrag{beta=1}[Bl][Bl][1]{$\beta = 1$}
\psfrag{beta=2}[Bl][Bl][1]{$\beta = 2$}
\psfrag{beta=10}[Bl][Bl][1]{$\beta = 10$}
\psfrag{beta=20}[Bl][Bl][1]{$\beta = 20$}
\psfrag{beta=50}[Bl][Bl][1]{$\beta = 50$}
\psfrag{beta=80}[Bl][Bl][1]{$\beta = 80$}
\psfrag{beta = 200}[Bl][Bl][1]{$\beta = 200$}
\psfrag{0}[Bl][Bl][1.0]{$0$}
\psfrag{1}[Bl][Bl][1.0]{$1$}
\psfrag{2}[Bl][Bl][1.0]{$2$}
\psfrag{3}[Bl][Bl][1.0]{$3$}
\psfrag{4}[Bl][Bl][1.0]{$4$}
\psfrag{-1}[Bl][Bl][1.0]{$-1$}
\psfrag{-2}[Bl][Bl][1.0]{$-2$}
\psfrag{-4}[Bl][Bl][1.0]{$-4$}
\psfrag{0.1}[Bl][Bl][1.0]{$\hspace{-0.1em}0.1$}
\psfrag{0.2}[Bl][Bl][1.0]{$\hspace{-0.1em}0.2$}
\psfrag{0.3}[Bl][Bl][1.0]{$\hspace{-0.1em}0.3$}
\psfrag{0.4}[Bl][Bl][1.0]{$\hspace{-0.1em}0.4$}
\psfrag{0.5}[Bl][Bl][1.0]{$\hspace{-0.1em}0.5$}
\psfrag{0.6}[Bl][Bl][1.0]{$\hspace{-0.1em}0.6$}
}
\label{fig:dot_spectra_U2}}
\subfloat[\strut]{
  \img{width=0.46\linewidth}{./dot_spectra_U4}
{\psfrag{om}[Bl][Bl][1.0]{$\omega$}
\psfrag{Aom}[Bl][Bl][1.0]{$A(\omega)$}
\psfrag{beta=0.1}[Bl][Bl][0.9]{$\beta = 0.1$}
\psfrag{beta=1}[Bl][Bl][0.9]{$\beta = 1$}
\psfrag{beta=5}[Bl][Bl][0.9]{$\beta = 5$}
\psfrag{beta=10}[Bl][Bl][0.9]{$\beta = 10$}
\psfrag{beta=20}[Bl][Bl][0.9]{$\beta = 20$}
\psfrag{beta=40}[Bl][Bl][0.9]{$\beta = 40$}
\psfrag{beta=80}[Bl][Bl][0.9]{$\beta = 80$}
\psfrag{beta=150}[Bl][Bl][0.9]{$\beta = 150$}
\psfrag{beta=300}[Bl][Bl][0.9]{$\beta = 300$}
\psfrag{0}[Bl][Bl][1.0]{$0$}
\psfrag{1}[Bl][Bl][1.0]{$1$}
\psfrag{2}[Bl][Bl][1.0]{$2$}
\psfrag{3}[Bl][Bl][1.0]{$3$}
\psfrag{4}[Bl][Bl][1.0]{$4$}
\psfrag{1}[Bl][Bl][1.0]{$-1$}
\psfrag{-2}[Bl][Bl][1.0]{$-2$}
\psfrag{-3}[Bl][Bl][1.0]{$-3$}
\psfrag{-4}[Bl][Bl][1.0]{$-4$}
\psfrag{0.1}[Bl][Bl][1.0]{$\hspace{-0.7em}0.1$}
\psfrag{0.2}[Bl][Bl][1.0]{$\hspace{-0.7em}0.2$}
\psfrag{0.3}[Bl][Bl][1.0]{$\hspace{-0.7em}0.3$}
\psfrag{0.4}[Bl][Bl][1.0]{$\hspace{-0.7em}0.4$}
\psfrag{0.5}[Bl][Bl][1.0]{$\hspace{-0.7em}0.5$}
\psfrag{0.6}[Bl][Bl][1.0]{$\hspace{-0.7em}0.6$}}
}
 \caption{(Color online) The spectral functions of the dot for (a) $U=2$ and (b) $U=4$, both at $V=0.5$. At $\beta = 0.1$ we see the high-temperature regime with weight centered around zero. Lowering the temperature we cross over into the local moment regime with two clearly seperated Hubbard bands at $\omega \approx \pm \frac{U}{2}$. Lowering the temperature further we see the emergence of the Kondo resonance at $\omega = 0$.}
 \label{fig:dot_spectra}
\end{figure}
To study the deflection of the edge state as well as the formation of the Kondo resonance, we also compute the site dependent single particle density  spectral function
\begin{equation}
 A_n(r,\omega) = -\frac{1}{\pi} \text{Im}\left(G_n(r, \omega + \I \eta)\right)
\end{equation}
for a small $\eta$ from the bath Green's function $G_n(r, z)$ at site $(r,n)$.
To obtain the spectral functions of the bath lattice we analytically continue the impurity self-energy to real frequencies and 
use Dyson's equation to access the lattice spectral functions as detailed in appendix \ref{sec:analytical_continuation}.
Additionally we get access to the term $C_n(r, \omega, \Sigma(\omega))$ of \autoref{eq:structurespectralfunction}.
\begin{figure}
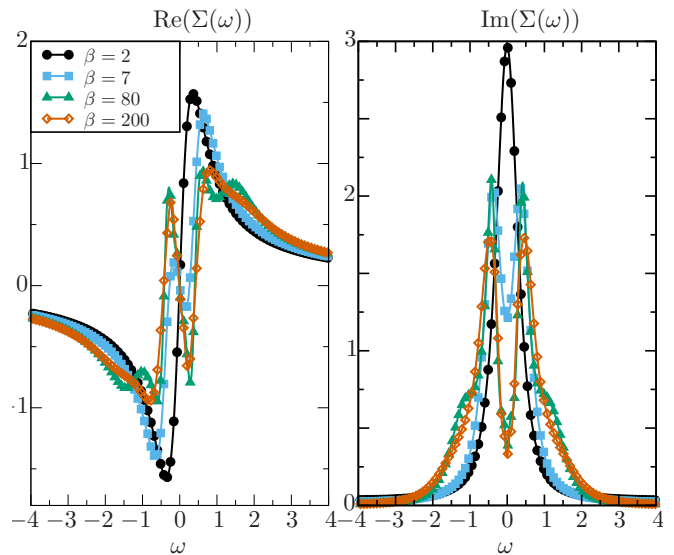

\img{width=\linewidth}{./selfenergy_overview}
{\psfrag{om}[Bl][Bl][1.9]{$\omega$}
\psfrag{Resigma}[Bl][Bl][1.9]{$\text{Re}(\Sigma(\omega))$}
\psfrag{Imsigma}[Bl][Bl][1.9]{$\text{Im}(\Sigma(\omega))$}
\psfrag{beta=0.1}[Bl][Bl][1.5]{$\beta = 0.1$}
\psfrag{beta=2}[Bl][Bl][1.5]{$\beta = 2$}
\psfrag{beta=7}[Bl][Bl][1.5]{$\beta = 7$}
\psfrag{beta=80}[Bl][Bl][1.5]{$\beta = 80$}
\psfrag{beta=200}[Bl][Bl][1.5]{$\beta = 200$}
\psfrag{0}[Bl][Bl][1.9]{$\hspace{-0.1em}0$}
\psfrag{1}[Bl][Bl][1.9]{$\hspace{-0.1em}1$}
\psfrag{2}[Bl][Bl][1.9]{$\hspace{-0.1em}2$}
\psfrag{3}[Bl][Bl][1.9]{$\hspace{-0.1em}3$}
\psfrag{0.5}[Bl][Bl][1.9]{}
\psfrag{1.5}[Bl][Bl][1.9]{}
\psfrag{2.5}[Bl][Bl][1.9]{}
\psfrag{-1}[Bl][Bl][1.9]{$\hspace{-0.7em}-1$}
\psfrag{-2}[Bl][Bl][1.9]{$\hspace{-0.7em}-2$}
\psfrag{-3}[Bl][Bl][1.9]{$\hspace{-0.7em}-3$}
\psfrag{-4}[Bl][Bl][1.9]{$\hspace{-0.7em}-4$}
\psfrag{4}[Bl][Bl][1.9]{$4$}
}
\caption{(Color online) Real and imaginary part of the self-energy $\Sigma(\omega)$ in different temperature regimes for $U=2$.
$\text{Im}(\Sigma(\omega))$ shows a simple lorentzian shape for $\beta = 2$ in the local moment regime.
Crossing over to the Kondo regime we see the development of a two peak structure. Since $\Sigma(z)$
is a holomorphic function its real and imaginary part are linked via the Kramers-Kronig relations.
}
\label{fig:selfenergies}
\end{figure}
\autoref{fig:selfenergies} shows the self-energy -- which is non-vanishing only on the impurity site -- in different temperature regimes.
Since $\Sigma(z)$ is a holomorphic function its real and imaginary part are linked via the Kramers-Kronig relations,
therefore we find the peak-structure-like features of the imaginary part as zero-crossings in the real part.
The apparent symmetry of the self-energies is due to the kernel we have used for the analytical continuation procedure,
starting from the imaginary part of the Green's function in Matsubara frequencies as input data.
Since we have for the quantity $\Sigma'$ that
\begin{equation}
 \Sigma' (i \omega_n) = \int \limits_{-\infty}^\infty d\omega \frac{\text{Im} \left(\Sigma ' (\omega)\right)}{i \omega_n - \omega}
\end{equation}
its imaginary part is
\begin{equation}
 \text{Im}\left( \Sigma ' (i \omega_n) \right) = - \omega_n \int \limits_{-\infty}^\infty d\omega \frac{\text{Im} \left(\Sigma ' (\omega)\right)}{i \omega_n^2 + \omega^2},
\end{equation}
which is symmetric in $\omega$. Note that $\Sigma$ and $\Sigma'$ are linked via a simple rescaling.
The temperature dependence of the self-energy documents the crossover from a single peak to a three-peak structure in the impurity spectral function shown in \autoref{fig:dot_spectra}.
At temperature scales above the Hubbard $U$, correlations effects are not important and the self-energy essentially vanishes such that the  impurity
spectral function reduces to the non-interacting one with a single central peak pinned at the Fermi energy due to particle-hole symmetry.
Lowering the temperature, we observe the formation of three zero crossings in $\text{Re} \left(\Sigma\right)$.
Two are roughly located at $\omega = \pm U/2$, are heavily damped since $\text{Im} \left(\Sigma\right)$ is large for those frequencies,
and correspond to the upper and lower Hubbard features.
The central zero crossing corresponds to the Kondo resonance.
It has a narrow line shape and hence has a small value of $\text{Im} \left(\Sigma\right)$.
That these central crossings have the same slope is somehow expected,
since the slope should be proportional to the Kondo temperature of the system \cite{Fakher_2004}.

\subsection{Estimating the Kondo temperature}
\begin{figure}
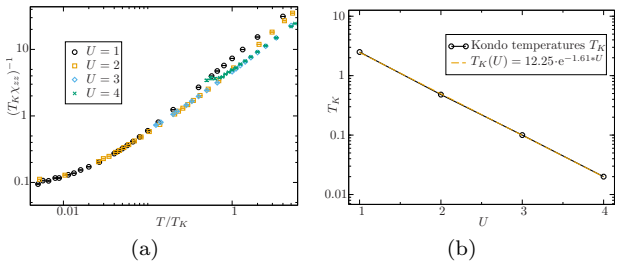

\hspace{-2em}
\subfloat[\strut]{
\img{width=0.45\linewidth}{./datacollapse}
{\psfrag{TTK}[Bl][Bl][1.0]{$T/T_K$}
\psfrag{UUU=1}[Bl][Bl][1.0]{$U=1$}
\psfrag{U=2}[Bl][Bl][1.0]{$U=2$}
\psfrag{U=3}[Bl][Bl][1.0]{$U=3$}
\psfrag{U=4}[Bl][Bl][1.0]{$U=4$}
\psfrag{10}[Bl][Bl][1.0]{$10$}
\psfrag{1}[Bl][Bl][1.0]{$1$}
\psfrag{invchi}[Bl][Bl][1.0]{$(T_K \chi_{zz})^{-1}$}
\psfrag{0.1}[Bl][Bl][1.0]{$0.1$}
\psfrag{0.01}[Bl][Bl][1.0]{$0.01$}
}\label{fig:datacollapse}
}
\subfloat[\strut]{
\img{width=0.45\linewidth}{./exponential_fit_of_TK}
{\psfrag{TK}[Bl][Bl][1.0]{$T_K$}
\psfrag{1}[Bl][Bl][1.0]{$1$}
\psfrag{2}[Bl][Bl][1.0]{$2$}
\psfrag{3}[Bl][Bl][1.0]{$3$}
\psfrag{4}[Bl][Bl][1.0]{$4$}
\psfrag{U}[Bl][Bl][1.0]{$U$}
\psfrag{TK(U)=12.247*Exp(-1.6054*U)}[Bl][Bl][1.0]{$T_K(U)=12.25\hspace{-0.4ex}\cdot \hspace{-0.4ex}\mathrm{e}^{-1.61*U}$}
\psfrag{0.1}[Bl][Bl][1.0]{$0.1$}
\psfrag{0.01}[Bl][Bl][1.0]{$0.01$}
\psfrag{10}[Bl][Bl][1.0]{$10$}
\psfrag{Kondo temperatures}[Bl][Bl][1.0]{Kondo temperatures $T_K$}
}\label{fig:expfit}
}
\caption{(Color online) The data for estimating the Kondo temperature $T_K$ for the parameters $V=0.5$ and $\lambda = 0.1$.
Since $\chi_{zz}$ is a universal function with $T_K$ as the only parameter we show in (a) a data collapse of the data for different values of $U$.
The data points roughly lie on the same function, with the $U=1$ points extending to the lowest relative temperatures.
(b) shows in a log-plot the obtained Kondo temperatures with its dependence on $U$. 
It shows an exponential dependence on the interaction $U$ that is expected for the symmetric Anderson model.
}
\end{figure}

To get an estimate of the involved Kondo temperatures $T_K$ we carried out a data collapse in \autoref{fig:datacollapse} 
of the susceptibilities shown in \autoref{fig:chi_zz}.
from \autoref{fig:datacollapse}.
It is known that the susceptibility of the Kondo model should be a universal function
\begin{equation}
 T_K \chi_{zz} = F\left( \frac{T}{T_K} \right)
\end{equation}
with $T_K$ as the only scaling parameter\cite{Hewson1997}.
We find the numerical values for the inverse Kondo temperature $\beta_K = T_K^{-1}$
\begin{center}
 \begin{tabular}{|c|c|c|c|c|}
  $U$ & 1 & 2 & 3 & 4 \\ \hline
  $\beta_K$ & 0.4 & 2.08 & 10 & 50 \\
 \end{tabular}
\end{center}
With the Kondo temperatures at hand we performed a cross check of the obtained values of $T_K$ and assumed the validity
of the asymptotic behaviour of the Kondo temperature for the symmetric Anderson model, given by
\begin{equation}
 T_K \propto e^{-\frac{U}{8 V^2 \rho_0}},
\end{equation}
with the density of states $\rho_0$\cite{Hewson1997}.
Note that for the Kane-Mele model in the considered parameter range, the  Fermi velocity of the edge state is set by the spin-orbit coupling.
Hence $\rho_0 \propto 1/\lambda$.
The straight line in the log-plot of \autoref{fig:expfit} indeed confirms this behaviour.
Additionally, another cross-check is available by means of the impurity spectral functions of \autoref{fig:dot_spectra}.
For $U=2$ we have an inverse Kondo temperature of $\beta_K \approx 2.1 $ which is consistent since somewhere in the range $\beta=2$ and $\beta=10$
the Kondo resonance starts to build-up at $\omega = 0$.
For $U=4$ we get $\beta_K = 50$ which is again consistent with the data for the spectral function.

\subsection{High-temperature regime}
The high-temperature regime is defined by the lack of any visible structure in the dot's spectral function.
$\beta = 0.1$ in \autoref{fig:dot_spectra_U2} is a good example, it shows just some lorentzian peak around $\omega = 0$.
With the presence of the small parameter $\beta U $ it is obvious that any interaction-induced correlation effects
to the spectral functions are negligible. Therefore in this regime the spectral function is that of the uncorrelated system.
Since all correlation effects are thermally washed out, the notion of a self-energy is meaningless and hence the self-energy contribution
$C_n(r,\omega,\Sigma(\omega))$ in \autoref{eq:structurespectralfunction} vanishes.
Therefore, the lattice spectral functions look indistinguishable to the non-interacting case \autoref{fig:fullspectra_imp_no_correlation}.
Since we are considering the particle-hole symmetric point, the occupancy of the impurity site is pinned to half-filling.
Hence in the absence of interactions the double occupancy $\langle n^d_\uparrow n^d_\downarrow \rangle$ takes the value $0.25$.
As apparent in \autoref{fig:docc_comparison} this value is approached as $\beta \rightarrow 0$.
Finally, in this high-temperature limit the spin susceptibility shows a $1/T$ behaviour as apparent from \autoref{fig:chi_zz}.

\subsection{Local Moment regime}
The formation of a local moment is at best characterized by the quantity 
\begin{equation}
\frac{\langle n^d_\uparrow n^d_\downarrow\rangle}{\langle n^d_\uparrow \rangle \langle n^d_\downarrow\rangle}.
\end{equation}
Since as mentioned previously the denominator of the above equation is pinned to $0.25$ by particle-hole symmetry,
the formation of the local moment boils down to the suppression of the double occupancy.
We will associate a characteristic energy scale of this regime by the dip in the double occupancy of \autoref{fig:docc_comparison}. 
The local moment regime corresponds to the regime where the Hubbard bands at $\omega \approx \frac{U}{2}$ develop in the dot's spectral function.
In \autoref{fig:dot_spectra_U2}, $\beta = 2$ is a good example of that. 
\begin{figure*}
\begin{tabular}{cc}
\subfloat[\strut]{%
 \def\svgwidth{0.55\linewidth}
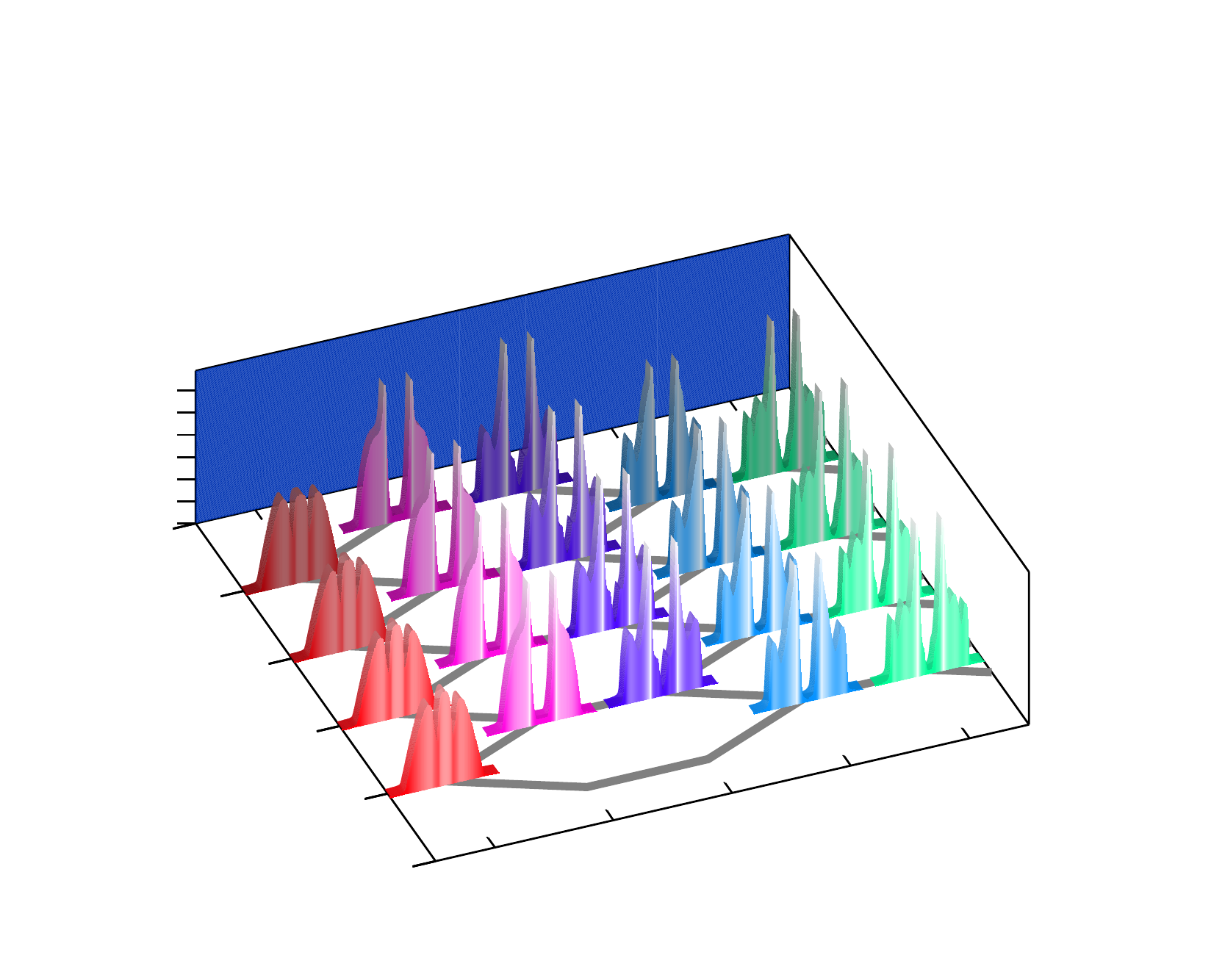
\label{fig:fullspectras_beta2}}&
\hspace{-0.04\linewidth}\subfloat[\strut]{%
\img{width=0.44\linewidth}{./fullspectra_beta2}
{\psfrag{om}[Bl][Bl][1.0]{$\omega$}
\psfrag{Aom}[Bl][Bl][1.0]{$A(\omega)$}
\psfrag{0}[Bl][Bl][1.0]{$0$}
\psfrag{1}[Bl][Bl][1.0]{$1$}
\psfrag{2}[Bl][Bl][1.0]{$2$}
\psfrag{3}[Bl][Bl][1.0]{$3$}
\psfrag{4}[Bl][Bl][1.0]{$4$}
\psfrag{5}[Bl][Bl][1.0]{$5$}
\psfrag{-1}[Bl][Bl][1.0]{$\hspace{-0.5em}-1$}
\psfrag{-2}[Bl][Bl][1.0]{$\hspace{-0.5em}-2$}
\psfrag{-3}[Bl][Bl][1.0]{$\hspace{-0.5em}-3$}
\psfrag{-4}[Bl][Bl][1.0]{$\hspace{-0.5em}-4$}
\psfrag{-5}[Bl][Bl][1.0]{$\hspace{-0.5em}-5$}
\psfrag{nnn = 0}[Bl][Bl][1.0]{$n=0$}
\psfrag{n = 1}[Bl][Bl][1.0]{$n=1$}
\psfrag{n = 2}[Bl][Bl][1.0]{$n=2$}
\psfrag{n = 3}[Bl][Bl][1.0]{$n=3$}
\psfrag{n = 4}[Bl][Bl][1.0]{$n=4$}
\psfrag{n = 5}[Bl][Bl][1.0]{$n=5$}
\psfrag{0.5}[Bl][Bl][1.0]{\hspace{-0.3em}$0.5$}
\psfrag{1.0}[Bl][Bl][1.0]{$1.0$}
\psfrag{1.5}[Bl][Bl][1.0]{\hspace{-0.3em}$1.5$}}
}\\
 \subfloat[\strut]{%
 \def\svgwidth{0.55\linewidth}
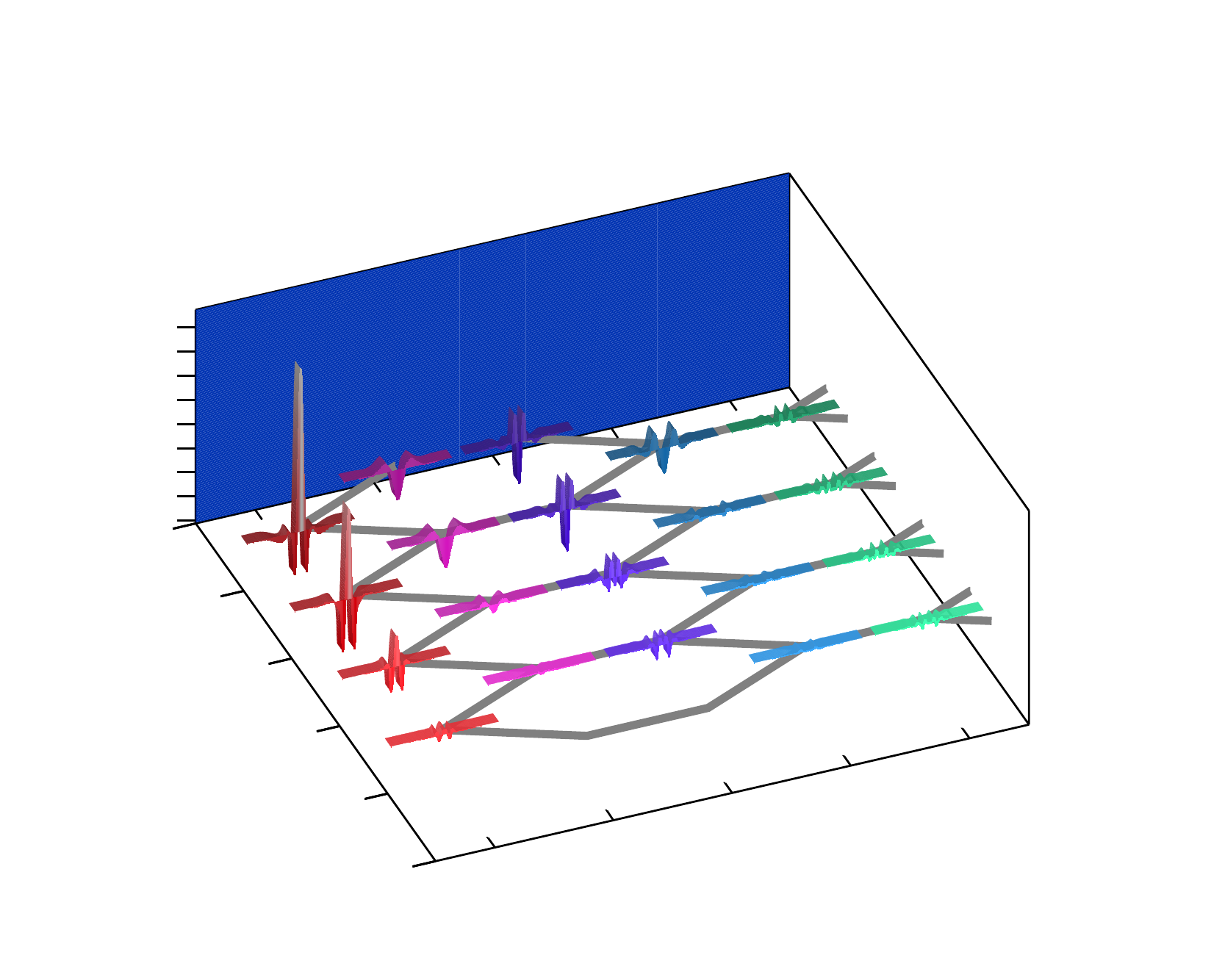
 }&
\hspace{-0.04\linewidth}\subfloat[\strut]{%
\img{width=0.44\linewidth}{./selfenergy_effect_beta2}
{\psfrag{om}[Bl][Bl][1.0]{$\omega$}
\psfrag{Aom}[Bl][Bl][1.0]{$C_n(0,\omega,\Sigma)$}
\psfrag{0}[Bl][Bl][1.0]{$0$}
\psfrag{1}[Bl][Bl][1.0]{$1$}
\psfrag{2}[Bl][Bl][1.0]{$2$}
\psfrag{3}[Bl][Bl][1.0]{$3$}
\psfrag{4}[Bl][Bl][1.0]{$4$}
\psfrag{-1}[Bl][Bl][1.0]{$\hspace{-0.5em}-1$}
\psfrag{-2}[Bl][Bl][1.0]{$\hspace{-0.5em}-2$}
\psfrag{-3}[Bl][Bl][1.0]{$\hspace{-0.5em}-3$}
\psfrag{-4}[Bl][Bl][1.0]{$\hspace{-0.5em}-4$}
\psfrag{r0,n= 0}[Bl][Bl][1.0]{$n=0$}
\psfrag{r0,n= 1}[Bl][Bl][1.0]{$n=1$}
\psfrag{r0,n= 2}[Bl][Bl][1.0]{$n=2$}
\psfrag{r0,n= 3}[Bl][Bl][1.0]{$n=3$}
\psfrag{r0,n= 4}[Bl][Bl][1.0]{$n=4$}
\psfrag{r0,n= 5}[Bl][Bl][1.0]{$n=5$}
\psfrag{0.2}[Bl][Bl][1.0]{\hspace{-0.3em}$0.2$}
\psfrag{-0.2}[Bl][Bl][1.0]{\hspace{-1em}$-0.2$}
\psfrag{0.4}[Bl][Bl][1.0]{\hspace{-0.3em}$0.4$}
\psfrag{0.6}[Bl][Bl][1.0]{\hspace{-0.3em}$0.6$}
\psfrag{-0.4}[Bl][Bl][1.0]{\hspace{-1em}$-0.4$}
\psfrag{-0.6}[Bl][Bl][1.0]{\hspace{-1em}$-0.6$}
\psfrag{-0.8}[Bl][Bl][1.0]{\hspace{-1em}$-0.8$}
}\label{fig:selfenergy_effect_beta2}}
\end{tabular}
\caption{(Color online) The spectra for the local moment regime at $\beta=2, U=2, V=0.5$ and $\lambda=0.1$. (a) shows all spectral functions around the
impurity, whereas (b) shows a cut of the spectral functions along $r=0$.
(c) and (d) show the self-energy contribution $C_n(r,\omega, \Sigma(\omega))$.
$C_n(r=0,\omega, \Sigma(\omega))$ is shown in (d).
A neat thing is that the contribution $C$ due to the self-energy seems to exactly cancel the effects
of $B$, the hybridization of the impurity with the bath, since the contribution in e.g. \autoref{fig:selfenergy_effect_beta2} has opposite sign of \autoref{fig:hyb_effect_cut}.}
\end{figure*}
At these intermediate temperatures we have one occupied spin state below the Fermi energy at $- \frac{U}{2}$ with a single electron.
This gives rise to essentially a free spin-$\frac{1}{2}$ degree of freedom: a local moment.
The energy scale at  which the double occupancy  is enhanced before saturating  marks the onset of the super exchange scale. 
This scale is set by $V^2/U$ for our particle-hole symmetric impurity problem.
In the context of the helical liquid the spin-flip scattering generated by the super-exchange scale corresponds to single particle back scattering.
These processes will hence reduce the conductance as noted in ref.\cite{2009PhRvL.102y6803M}. 
In the local moment domain, the site resolved single particle spectral function of the edge and bulk states 
in \autoref{fig:fullspectras_beta2} show no sign of Kondo screening.
In particular there is no deflection of the edge current perceivable.
\autoref{fig:change_beta2} shows that the change of $A_n(r,\omega)$ relative to some very distant 
point of reference has a very small amplitude.
Due to \autoref{eq:structurespectralfunction} this means an approximate cancellation of the hybridisation effect from 
\autoref{fig:hyb_effect_cut} with the now present self-energy effect shown
in
\autoref{fig:selfenergy_effect_beta2}.
The edge state is in that sense restored to the true boundary of the lattice as if no impurity were present.
This confirms the robustness of the edge state to a free local moment at least in view of the spectral functions.
Nevertheless we expect the conductance through this edge state to decrease in this regime due to the possible backscattering spin-flip processes
which is now available due to the impurity\cite{2013arXiv1301.6501H}.
Lowering the temperature further should break this match and deform the edge state again,
with the emergence of the Kondo resonance at $\omega = 0$ in the impurity's spectral function.
\subsection{Kondo regime}
\begin{figure*}
\begin{tabular}{cc}
\subfloat[\strut]{
  \def\svgwidth{0.55\linewidth}
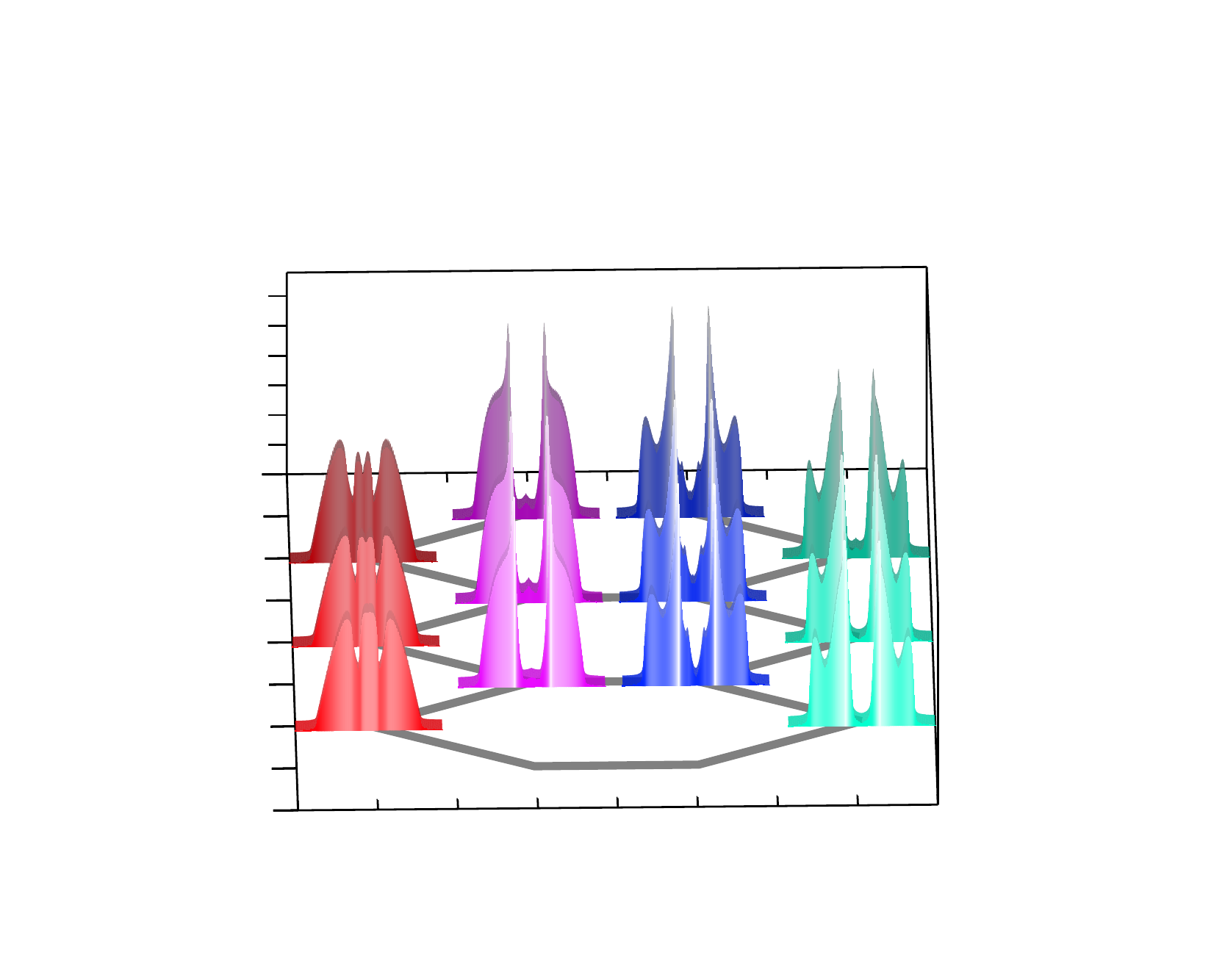}
&\hspace{-0.04\linewidth}
\subfloat[\strut]{
\img{width=0.44\linewidth}{./full_spectra_beta200_U2}
{\psfrag{om}[Bl][Bl][1.0]{$\omega$}
\psfrag{Aom}[Bl][Bl][1.0]{$A(\omega)$}
\psfrag{0}[Bl][Bl][1.0]{$0$}
\psfrag{1}[Bl][Bl][1.0]{$1$}
\psfrag{2}[Bl][Bl][1.0]{$2$}
\psfrag{3}[Bl][Bl][1.0]{$3$}
\psfrag{4}[Bl][Bl][1.0]{$4$}
\psfrag{5}[Bl][Bl][1.0]{$5$}
\psfrag{-1}[Bl][Bl][1.0]{$\hspace{-0.5em}-1$}
\psfrag{-2}[Bl][Bl][1.0]{$\hspace{-0.5em}-2$}
\psfrag{-3}[Bl][Bl][1.0]{$\hspace{-0.5em}-3$}
\psfrag{-4}[Bl][Bl][1.0]{$\hspace{-0.5em}-4$}
\psfrag{r = 0, n = 0}[Bl][Bl][1.0]{$n=0$}
\psfrag{r = 0, n = 1}[Bl][Bl][1.0]{$n=1$}
\psfrag{r = 0, n = 2}[Bl][Bl][1.0]{$n=2$}
\psfrag{r = 0, n = 3}[Bl][Bl][1.0]{$n=3$}
\psfrag{r = 0, n = 4}[Bl][Bl][1.0]{$n=4$}
\psfrag{r = 0, n = 5}[Bl][Bl][1.0]{$n=5$}
\psfrag{0.5}[Bl][Bl][1.0]{$\hspace{-0.5em}0.5$}
\psfrag{1.0}[Bl][Bl][1.0]{$\hspace{-0.1em}1.0$}
\psfrag{1.5}[Bl][Bl][1.0]{$\hspace{-0.5em}1.5$}}
\label{fig:full_spectra_beta200_U2_cut}
}\\
\subfloat[\strut]{
  \def\svgwidth{0.55\linewidth}
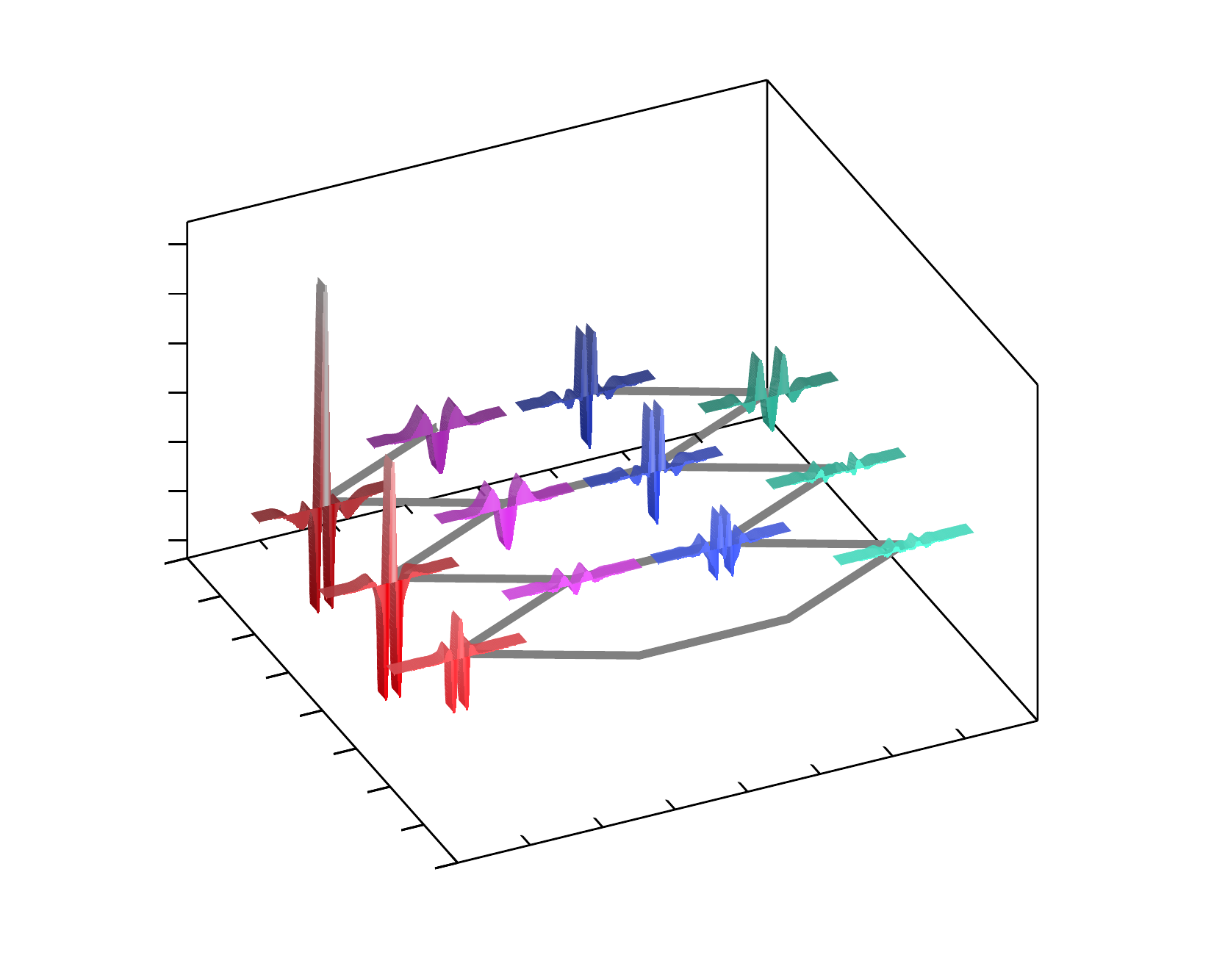}
&\hspace{-0.04\linewidth}\subfloat[\strut]{
\img{width=0.44\linewidth}{./selfenergy_effect_beta200_U2}
{\psfrag{om}[Bl][Bl][1.0]{$\omega$}
\psfrag{Aom}[Bl][Bl][1.0]{$C_n(0,\omega,\Sigma)$}
\psfrag{0}[Bl][Bl][1.0]{$0$}
\psfrag{1}[Bl][Bl][1.0]{$1$}
\psfrag{2}[Bl][Bl][1.0]{$2$}
\psfrag{3}[Bl][Bl][1.0]{$3$}
\psfrag{4}[Bl][Bl][1.0]{$4$}
\psfrag{-1}[Bl][Bl][1.0]{$\hspace{-0.5em}-1$}
\psfrag{-2}[Bl][Bl][1.0]{$\hspace{-0.5em}-2$}
\psfrag{-3}[Bl][Bl][1.0]{$\hspace{-0.5em}-3$}
\psfrag{-4}[Bl][Bl][1.0]{$\hspace{-0.5em}-4$}
\psfrag{r = 0, n = 0}[Bl][Bl][1.0]{$n=0$}
\psfrag{r = 0, n = 1}[Bl][Bl][1.0]{$n=1$}
\psfrag{r = 0, n = 2}[Bl][Bl][1.0]{$n=2$}
\psfrag{r = 0, n = 3}[Bl][Bl][1.0]{$n=3$}
\psfrag{r = 0, n = 4}[Bl][Bl][1.0]{$n=4$}
\psfrag{r = 0, n = 5}[Bl][Bl][1.0]{$n=5$}
\psfrag{0.2}[Bl][Bl][1.0]{$\hspace{-0.5em}0.2$}
\psfrag{-0.2}[Bl][Bl][1.0]{$\hspace{-1em}-0.2$}
\psfrag{0.4}[Bl][Bl][1.0]{$\hspace{-0.5em}0.4$}
\psfrag{0.6}[Bl][Bl][1.0]{$0.6$}
\psfrag{-0.4}[Bl][Bl][1.0]{$-0.4$}
\psfrag{-0.6}[Bl][Bl][1.0]{$-0.6$}
\psfrag{-0.8}[Bl][Bl][1.0]{$-0.8$}}\label{fig:selfenergy_effects_beta200}}
\end{tabular}
\caption{(Color online) For the parameters $\beta=200, U=2, V=0.5$ and $\lambda=0.1$, (a) shows a frontal view on the full spectral functions. The dip at the impurity is visible as well as the progression to the full edge state
at the edge. (b) shows a cut of the same spectral functions now for increasing $n$. The displacement of the edge into the bulk is visible.  
(c) and (d) show the effect due to the self-energy $C_n(r, \omega,\Sigma(\omega))$.
Especially figure (d) shows that in comparison with \autoref{fig:selfenergy_effect_beta2} the functional form of $C$ seems to be the same although we have
increased $\beta$ by two magnitudes, but the amplitude is reduced from about $0.6$ to $0.4$.
}
\end{figure*}

Lowering the temperature further we enter the Kondo regime.
The prominent signature of this regime is the emergence of the Kondo resonance in the dot's spectral functions as  seen for example in \autoref{fig:dot_spectra_U2} at $\beta = 200$.  
The Kondo resonance shows up as a dip in the local spectral function of the edge state electrons at the position of the impurity ($r=0$ and $n=0$)
as seen in \autoref{fig:full_spectra_beta200_U2_cut}.
The origin of these spectral features lies in the formation of the Kondo singlet which entangles the impurity spin with the spins of the surrounding electrons.
On an energy scale set by the Kondo temperature we expect the impurity site to act as a potential scatterer and hence lead to the deflection of the edge current into the bulk.
As apparent, the dip in the spectral weight at $n = 0$ in \autoref{fig:selfenergy_effects_beta200}(b) is accompanied by the emergence of a {\it  peak} at $n = 1$ and $n = 2$. 
At this very low temperature scale, $T/T_K \simeq 0.01$, it is this deflection of the edge state discussed by Maciejko et al.\cite{2009PhRvL.102y6803M}  which restores the conductance to unitarity in the limit $T  \rightarrow 0$.
Although we have decreased the temperature by two magnitudes the form of the change in \autoref{fig:selfenergy_effects_beta200} still looks the same,
only its magnitude has decreased. 
But of course now the self-energy term $C$ and the hybridization term $B$ do not match anymore as nicely as for $\beta=2$.
Hence there is not a full cancellation of the hybridization effect.
A further decrease of the temperature should bring us deeper into the Kondo regime with a deflection of the edge state around the impurity as predicted by Maciejko et al. \cite{2009PhRvL.102y6803M}. 
This confirms that, to the outside world, the singlet state of a magnetic impurity and bath electrons has the same low-energy features as plain potential scattering.
We only expect this circumvention of the edge state on an energy scale set by the Kondo temperature $T_K$, since beyond this energy scale the equivalence to a potential scatterer is not tenable \cite{Noziere74}.
Note that since the width of the Kondo resonance is of the order of $T_K$, the dip in the bulk spectral functions is of the same size.
On the other hand, the width of the edge state is given by the spin-orbit coupling $\lambda > T_K$.
\begin{figure}
\begin{tabular}{cc}
 \subfloat[\strut]{
\def\svgwidth{0.55\linewidth}
 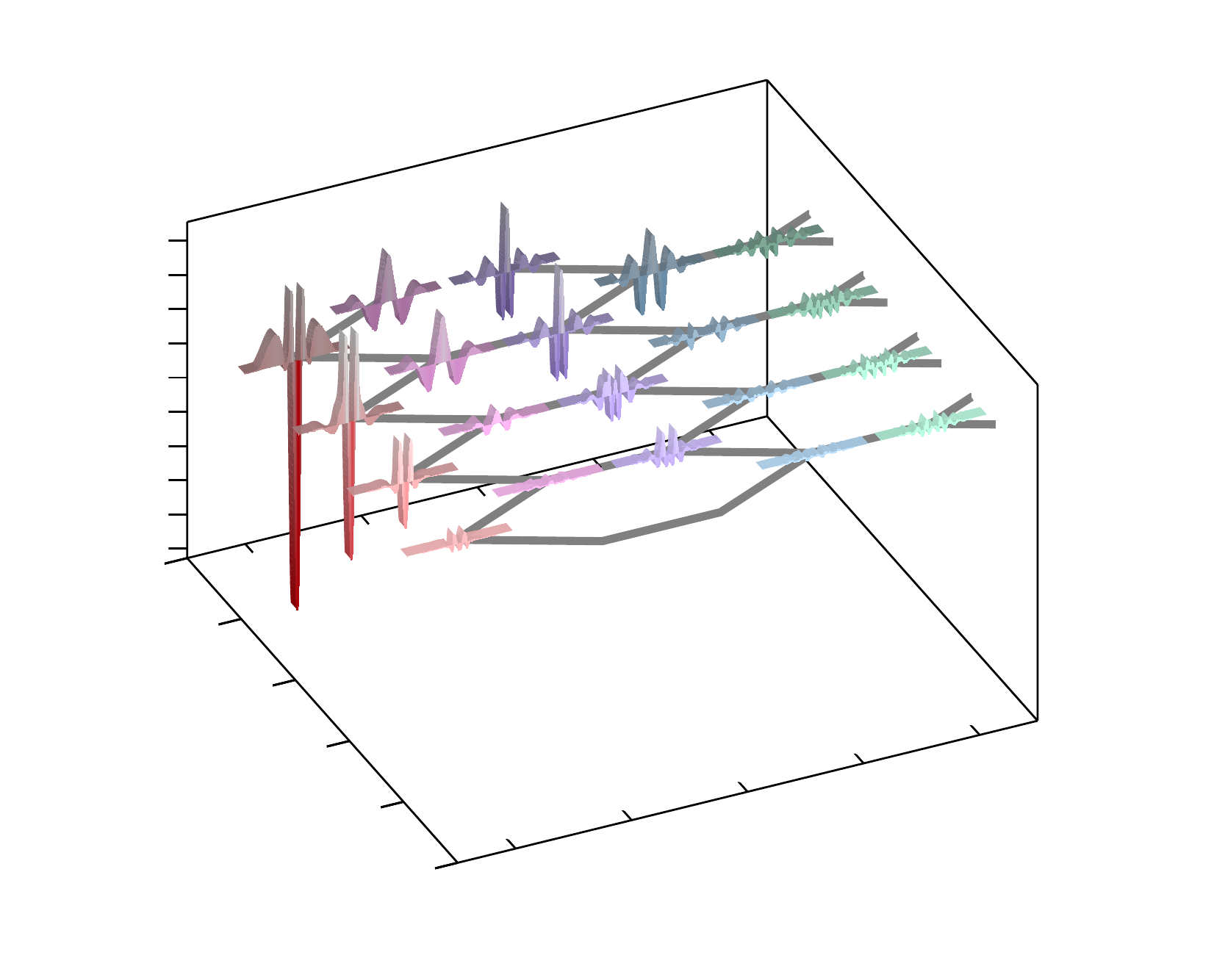\label{fig:change_nocorr}}&
\subfloat[\strut]{
\hspace{-3em}
 \def\svgwidth{0.55\linewidth}
 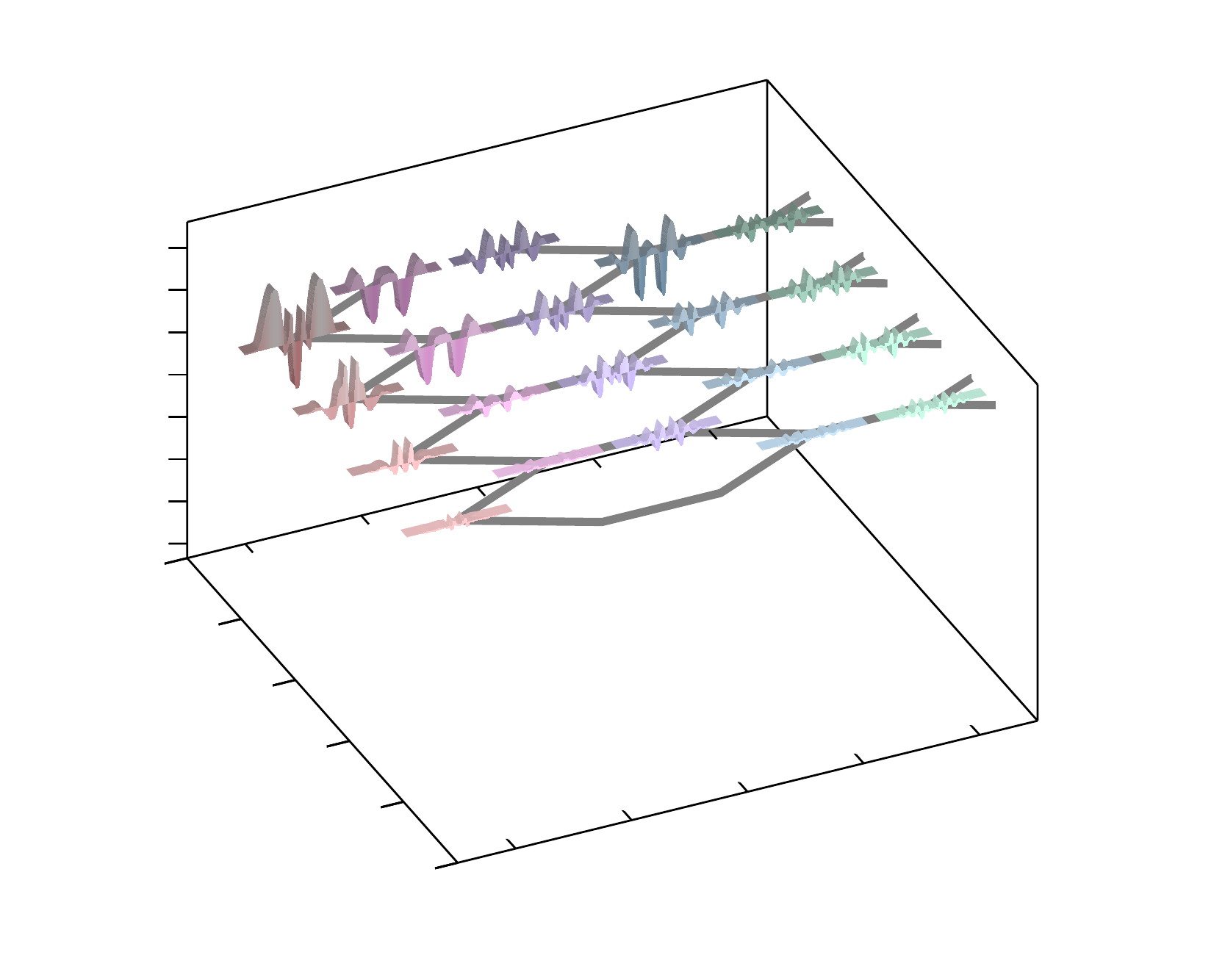\label{fig:change_beta2}}\\
 \subfloat[\strut]{
 \def\svgwidth{0.55\linewidth}
 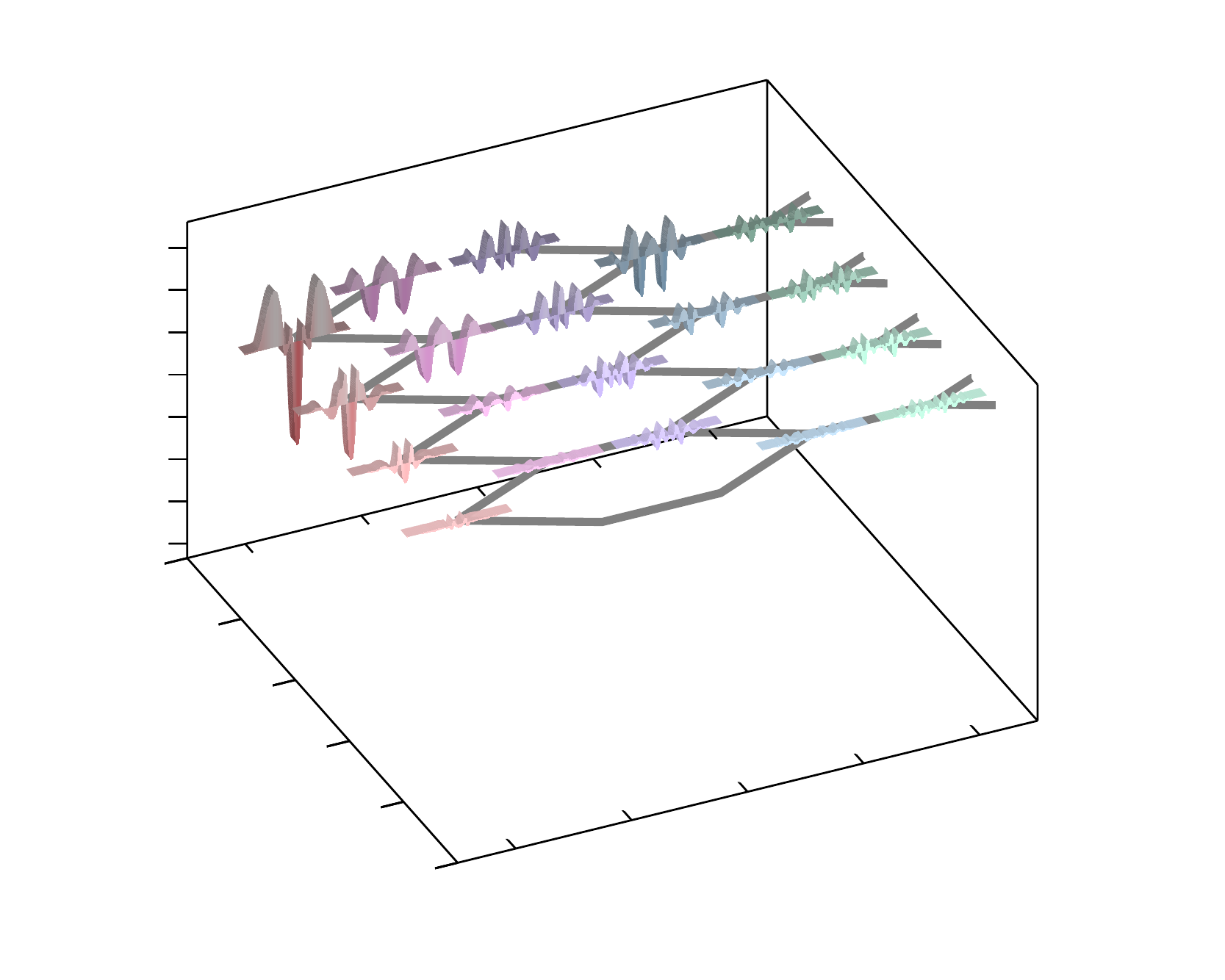\label{fig:change_beta7}} &
\subfloat[\strut]{
\hspace{-3em}
 \def\svgwidth{0.55\linewidth}
 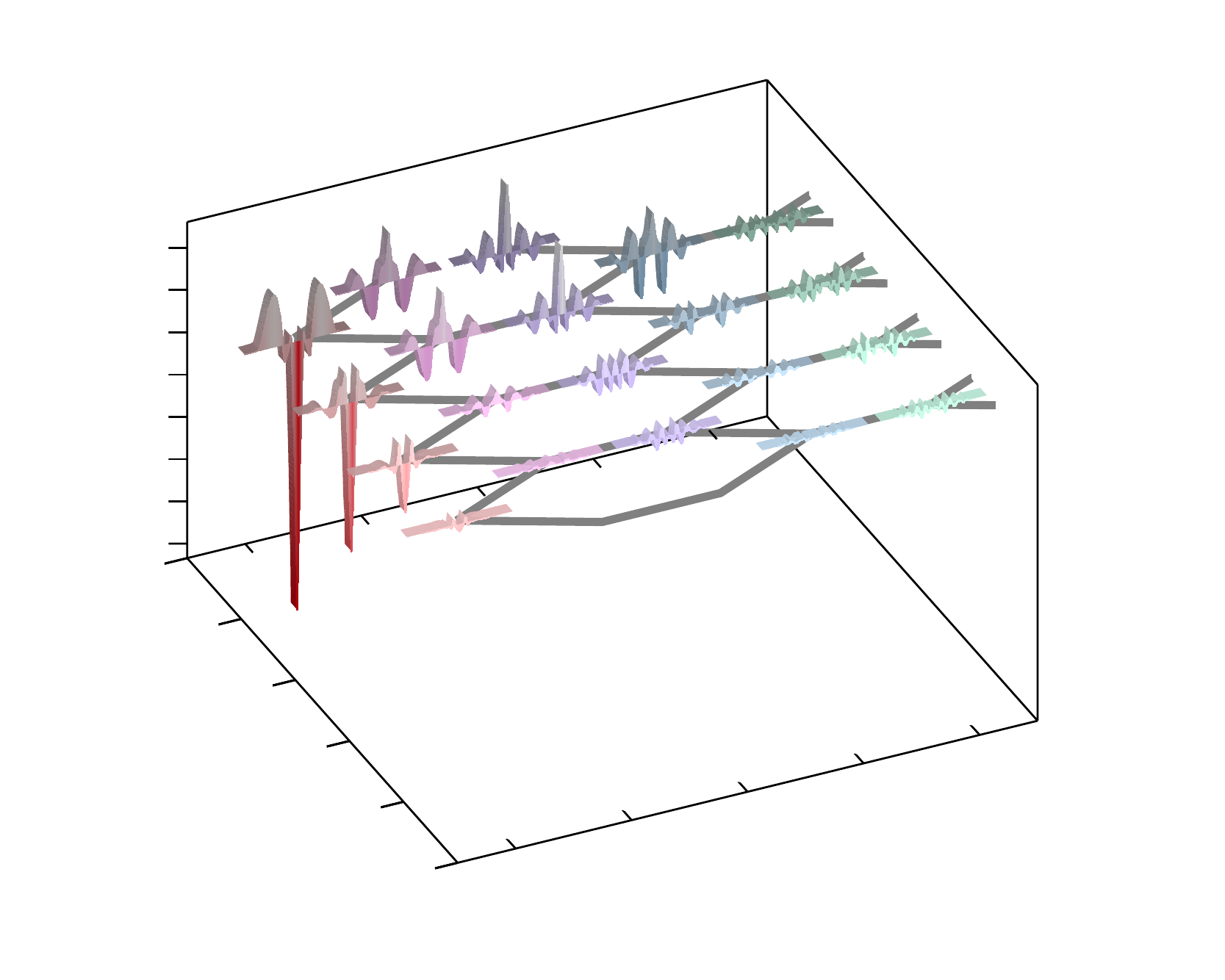\label{fig:change_beta200}}
\end{tabular}
\caption{Here we show the changes to the spectral functions in the vicinity of the impurity relative to some very distant 
reference point that feels no effect of the impurity. This corresponds to the quantity $\Delta_n(r,\omega) = A_n(r,\omega) - A^0_n(\omega)$.
(a) shows the changes due to an uncorrelated impurity.
(b) is the change in the local moment regime at $\beta = 2$. Note the small amplitude of $\Delta$, which means that in this regime the bath feels only 
a negligible effect of the impurity.
(c) is in the Kondo regime for $\beta = 7$.
(d) is in the Kondo regime for $\beta = 200$.
Note that only (b) - (d) share the same scale on the $\Delta_n(r,\omega)$ axis.
}\label{fig:change_with_reference}
\end{figure}
\autoref{fig:change_with_reference} shows the change attributable to the impurity measured against a very distant point in the same {\it orbital} $n$: 
$\Delta_n(r,\omega) = A_n(r,\omega) - A^0_n(\omega)$. 
We see the strong local effect of a potential scatterer in \autoref{fig:change_nocorr}.
In \autoref{fig:change_beta2} we see that a correlated impurity at $\beta = 2$ has a negligible effect on the bath. This corresponds to the local moment regime. As the 
temperature drops below the Kondo scale (see the data sets at  $\beta = 7$ in \autoref{fig:change_beta7} and at $\beta=200$ in \autoref{fig:change_beta200})
the same deflection of the edge current as observed for the potential scatterer emerges.

\section{Spatially resolved dot bulk spin spin correlation functions}
To provide a different point of view on our study of the Kondo cloud in the bath system that does not rely on an analytic continuation
procedure we now turn our attention towards the site resolved spin spin correlation functions
\begin{equation}
 \langle S^z_d S^z_c(r,n) \rangle
\end{equation}
between the impurity spin
$S^z_d = \frac{1}{2}\left(n^d_\uparrow - n^d_\downarrow \right) $
and a spin located at a particular site $S^z_c(r,n) = \frac{1}{2}\left( n_{r,n,\uparrow} - n_{r,n,\downarrow} \right)$ of a conduction electron. 
This enables us to define the Kondo cloud as the region of substantial entanglement of the impurity 
spin with a particular bath site. This is complementary to our previous results where we have defined the Kondo cloud
as the region where the edge state is suppressed.

\subsection{A 2D overview}
First, we consider the non-interacting case and find non-negligible correlations that are confined to the edge of the system.
Along the edge the spin spin correlations decay as $r^{-2}$ as already pointed out in Ref.\cite{PhysRevB.75.041307, Ishii_1978} for a 1D system of electrons.
Clearly, this power law holds only in the zero temperature limit and at finite temperature an exponential decay sets in beyond the thermal length scale  
$\xi_T \propto v_F \beta$ ($v_F$ is the Fermi velocity).
This similarity to the one-dimensional case provides yet another confirmation of the 1D nature of the edge state. 

The local moment regime just shows a small amount of correlation in
\autoref{fig:exponential_Kondo} since due to the high temperature all long-range effects are destroyed.
This is in contrast to the behaviour of the spectral functions in \autoref{fig:fullspectras_beta2} 
where we see no signature of the impurity in the bath system.
\begin{figure}
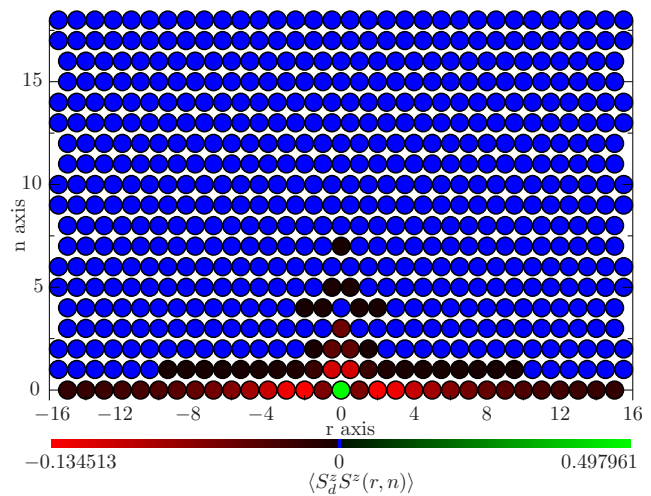

  \img{width=\linewidth}{./KondoCloud_beta100_Nx512_U2_lambda01}
 {
\psfrag{0}[Bl][Bl][1.6]{$\hspace{-0.4em}0$}
\psfrag{5}[Bl][Bl][1.6]{$5$}
\psfrag{10}[Bl][Bl][1.6]{$10$}
\psfrag{15}[Bl][Bl][1.6]{$15$}
\psfrag{n axis}[Bl][Bl][1.6]{n axis}
\psfrag{r axis}[Bl][Bl][1.6]{r axis}
\psfrag{240}[Bl][Bl][1.6]{$-16$}
\psfrag{244}[Bl][Bl][1.6]{$-12$}
\psfrag{248}[Bl][Bl][1.6]{$-8$}
\psfrag{252}[Bl][Bl][1.6]{$-4$}
\psfrag{256}[Bl][Bl][1.6]{$0$}
\psfrag{260}[Bl][Bl][1.6]{$4$}
\psfrag{264}[Bl][Bl][1.6]{$8$}
\psfrag{268}[Bl][Bl][1.6]{$12$}
\psfrag{272}[Bl][Bl][1.6]{$16$}
\psfrag{-0.134513}[Bl][Bl][1.6]{$-0.134513$}
\psfrag{0.497961}[Bl][Bl][1.6]{$0.497961$}
\psfrag{colorcoding}[Bl][Bl][1.6]{$\langle S^z_d S^z_c(r,n) \rangle$}
}
\caption{(Color online) A 2D overview of the color-coded correlation function $\langle S^z_d S^z_c(r,n) \rangle$ with the full spatial dependence at $\beta = 100, \lambda = 0.1$ and $U=2$.
The impurity is the green spot at $(r,n) = (0, 0)$ in the middle of the lower edge. Positive values are shades of green,
negative values are shades of red and neutral values are blue.
The color coding is linear. Note that this does not mean a linear perception of the color values.
The comparatively long chain of sites having various shades of red at the bottom of the diagram is the extent that
the correlation reaches into the edge state.
The correlation extends comparatively far into the edge state but is almost immediately suppressed away from the edge state.
}
\label{fig:spinspin_Kondo}
\end{figure}
If we now lower the temperature into the Kondo regime in \autoref{fig:spinspin_Kondo} we see that the effect of the impurity mostly extends into the helical liquid in the lower edge and develops some spatial structure.

\subsection{Correlation functions along the edge}
\begin{figure}
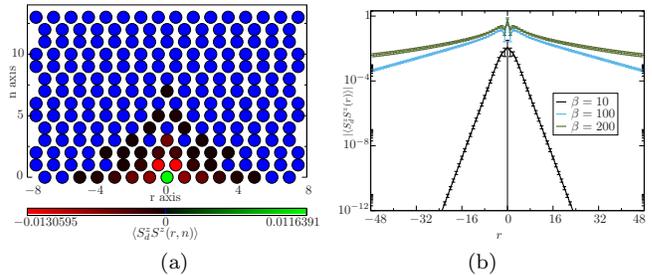

\subfloat[\strut]{
\hspace{-2em}
\img{width=0.48\linewidth}{./KondoCloud_beta10_Nx128_U2_lambda01}{
\psfrag{0}[Bl][Bl][1.0]{$\hspace{-0.4em}0$}
\psfrag{5}[Bl][Bl][1.0]{$5$}
\psfrag{10}[Bl][Bl][1.0]{$10$}
\psfrag{15}[Bl][Bl][1.0]{$15$}
\psfrag{n axis}[Bl][Bl][1.0]{n axis}
\psfrag{r axis}[Bl][Bl][1.0]{r axis}
\psfrag{-0.0130595}[Bl][Bl][1.0]{$-0.0130595$}
\psfrag{0.0116391}[Bl][Bl][1.0]{$0.0116391$}
\psfrag{-8}[Bl][Bl][1.0]{$-8$}
\psfrag{-4}[Bl][Bl][1.0]{$-4$}
\psfrag{8}[Bl][Bl][1.0]{$8$}
\psfrag{4}[Bl][Bl][1.0]{$4$}
\psfrag{spinspin}[Bl][Bl][1.0]{$\langle S^z_d S_c^z(r,n) \rangle$}
}}
\subfloat[\strut]{
 \img{width=0.48\linewidth}{./exponential_Kondo}
 {
 \psfrag{-64}[Bl][Bl][1.0]{$-64$}
 \psfrag{64}[Bl][Bl][1.0]{$64$}
 \psfrag{-48}[Bl][Bl][1.0]{$-48$}
 \psfrag{48}[Bl][Bl][1.0]{$\hspace{-0.5em}48$}
 \psfrag{-32}[Bl][Bl][1.0]{$-32$}
 \psfrag{32}[Bl][Bl][1.0]{$32$}
 \psfrag{-16}[Bl][Bl][1.0]{$-16$}
 \psfrag{16}[Bl][Bl][1.0]{$16$}
 \psfrag{0}[Bl][Bl][1.0]{$0$}
 \psfrag{1}[Bl][Bl][1.0]{$1$}
 \psfrag{0.0001}[Bl][Bl][1.0]{$10^{-4}$}
 \psfrag{1e-08}[Bl][Bl][1.0]{$\hspace{-0.4em}10^{-8}$}
 \psfrag{1e-12}[Bl][Bl][1.0]{$\hspace{-0.8em}10^{-12}$}
 \psfrag{beta = 10}[Bl][Bl][1.0]{$\beta = 10$}
 \psfrag{beta = 100}[Bl][Bl][1.0]{$\beta = 100$}
 \psfrag{beta = 200}[Bl][Bl][1.0]{$\beta = 200$}
 \psfrag{r axis}[Bl][Bl][1.0]{$r$}
 \psfrag{spinspin}[Bl][Bl][0.8]{$\vspace{-1em}\left| \langle S^z_d S_c^z(r)\rangle \right|$}
 }}
 \caption{(Color online) (a) shows the spin spin correlation functions in the local moment regime. Due to the rather high temperature we see that the correlation effects are quickly suppressed. Additionally, the absolute value of the correlations is an order of magnitude lower.
 The right figure, (b), provides a more detailed view of $\left| \langle S^z_d S_c^z(r,0) \rangle \right|$ at $U = 2$, but at different values of $\beta$
 in a logarithmic plot. Here we can notice the mirror symmetry around the impurity.
 We see that for $\beta = 10$, which is in the local moment regime (compare the spectral functions \autoref{fig:dot_spectra_U2}), the decay of the 
 correlation function is immediately exponential. This is consistent with the quick suppression in the left figure.
 Decreasing the temperature to $\beta = 100$ we see that the exponential decay sets in at  around $\xi_T \approx 8$ whereas for $\beta = 200$
 at $\xi_T \approx 16$.
 }
 \label{fig:exponential_Kondo}
\end{figure}

We now focus our attention on the correlations along the edge as a function of temperature, Hubbard interaction and spin-orbit coupling $\lambda$.
Borda et al.\ \cite{PhysRevB.75.041307} have studied the spatial behaviour of the spin spin correlation functions
of an Anderson impurity embedded in a one-dimensional wire as bath system.
They observed at a distance of $\xi_K \approx v_F \beta_K$ a crossover from an 
$r^{-1}$ behaviour to an $r^{-2}$ behaviour. For finite temperatures they predict the onset of an exponential
decay at $\xi_T \approx v_F \beta$. Their study shows that the spatial decay is oscillating with a wave vector
$k \propto k_F = \pi$. It can already be guessed from the 2D overviews that our system does not show oscillations,
which is consistent since $\cos(2 k_F) = 1$.
\begin{figure}
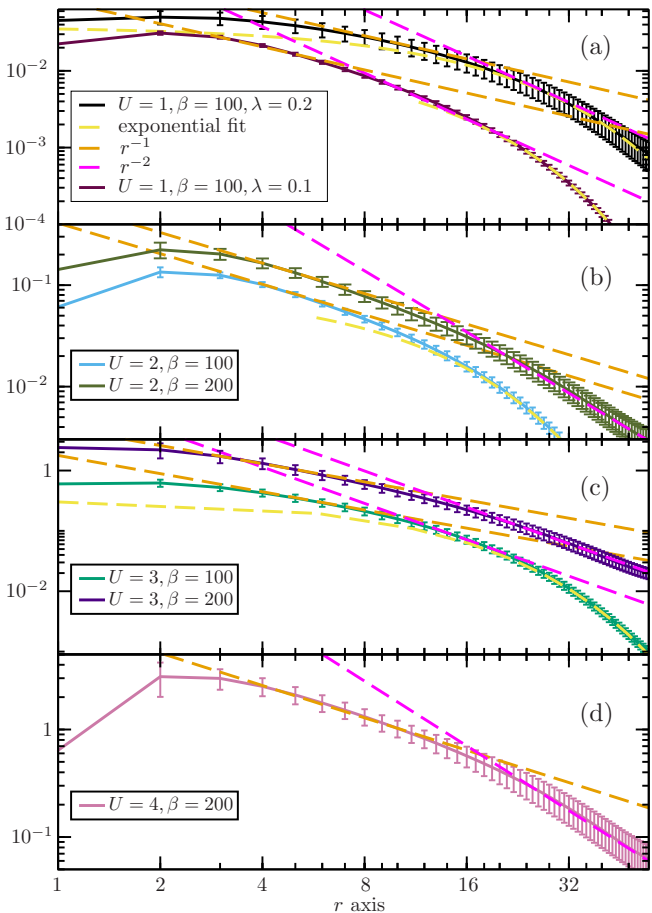

 \img{width=\linewidth}{./Kondo_edge}
 {
\psfrag{1}[Bl][Bl][1.6]{$\hspace{-0.2em}1$}
\psfrag{2}[Bl][Bl][1.6]{$2$}
\psfrag{4}[Bl][Bl][1.6]{$4$}
\psfrag{8}[Bl][Bl][1.6]{$8$}
\psfrag{16}[Bl][Bl][1.6]{$16$}
\psfrag{32}[Bl][Bl][1.6]{$32$}
\psfrag{64}[Bl][Bl][1.6]{$64$}
\psfrag{r axis}[Bl][Bl][1.6]{$r$ axis}
\psfrag{(a)}[Bl][Bl][2.0]{(a)}
\psfrag{(b)}[Bl][Bl][2.0]{(b)}
\psfrag{(c)}[Bl][Bl][2.0]{(c)}
\psfrag{(d)}[Bl][Bl][2.0]{(d)}
\psfrag{0.1}[Bl][Bl][1.6]{$\hspace{-1em}10^{-1}$}
\psfrag{0.01}[Bl][Bl][1.6]{$\hspace{-0.8em}10^{-2}$}
\psfrag{0.001}[Bl][Bl][1.6]{$\hspace{-0.4em}10^{-3}$}
\psfrag{0.0001}[Bl][Bl][1.6]{$10^{-4}$}
\psfrag{exponentialfit}[Bl][Bl][1.6]{exponential fit}
\psfrag{1/r}[Bl][Bl][1.6]{$r^{-1}$}
\psfrag{1/rr}[Bl][Bl][1.6]{$r^{-2}$}
\psfrag{U=1, beta=100,lambda=0.2}[Bl][Bl][1.5]{$U = 1, \beta=100, \lambda = 0.2$}
\psfrag{U=1, beta=100, lambda=0.1}[Bl][Bl][1.5]{$U = 1, \beta=100, \lambda = 0.1$}
\psfrag{U=2,\ \ beta=100}[Bl][Bl][1.5]{$\hspace{-0.5em}U = 2, \beta=100$}
\psfrag{U=2,\ \ beta=200}[Bl][Bl][1.5]{$\hspace{-0.5em}U = 2, \beta=200$}
\psfrag{U=3,\ \ beta=100}[Bl][Bl][1.5]{$\hspace{-0.5em}U = 3, \beta=100$}
\psfrag{U=3,\ \ beta=200}[Bl][Bl][1.5]{$\hspace{-0.5em}U = 3, \beta=200$}
\psfrag{U=4,\ \ beta=200}[Bl][Bl][1.5]{$\hspace{-0.5em}U = 4, \beta=200$}
 }
 \caption{(Color online) The spatial dependence of the correlation between the impurity's spin and the spin of a site of the edge
 $\left| \langle S^z_d S^z_c(r,n) \rangle \right|$ with a hybridization of $V=0.5$.
 Diagram (a) shows the dependence on the Fermi velocity at $U=1$ of  the cross-over point from an $r^{-1}$ to an $r^{-2}$ 
 behaviour, $\xi_K$, as well as of the point $\xi_T$ where the decay crosses over into an exponential law.
Note that for $\xi_K = v_F \beta_K$ both the Fermi velocity and Kondo temperature depend on $\lambda$: $v_F \propto \lambda$ and 
$\beta_K \propto \exp \left(\frac{\lambda}{J}\right)$.
%
 In the graph for $\lambda=0.2$ the exponential falloff is shifted outside of the visible lattice although the temperature is kept constant.
 The dashed lines are guides to the eye and are explained in the main text.
 Diagrams (b) - (d) are at $\lambda = 0.1$.
 In diagram (b) we show data for higher $T_K$ at $U=2$. We see that for the higher temperature $\beta=200$ the thermal decay at the end of the plot 
 is shifted outside of the visible part of the lattice and we can clearly identify the regimes with power law like behaviour.
 In the diagram for $\beta=100$ the $r^{-1}$ decay is already dominated by the thermal decay.
 Figure (c) and (d) show that further increments of $U$, which gives higher values of $\beta_K$, 
 shifts the cross-over point to larger distances.
 }
 \label{fig:Kondo_edge}
\end{figure}
\autoref{fig:Kondo_edge} shows that the general trend of these predictions made for a 1D chain of electrons also holds for the 
1D helical liquid of the edge state of a topological insulator if we perform an analysis similar to Ref. \cite{2012arXiv1207.7081P}.
In \autoref{fig:Kondo_edge}, (a) we find the dependence on the Fermi velocity, which is expected to be proportional to the spin-orbit coupling $\lambda$.
In the plot for $\lambda=0.2$ the expontial falloff is shifted outside of the visible part of the edge channel, although the plot for $\lambda = 0.1$ and $\lambda=0.2$ are at the same temperature.
In this plot we already have introduced a couple of lines that are meant as guides to the eye.
The yellow dashed lines denote the exponential decay that sets in beyond the thermal length scale $\xi_T \propto v_F \beta$.
The straight dashed magenta lines denote the power law decay $r^{-2}$ that is present at distances $\xi_K > r > \xi_T$.
Finally, the straight dashed orange lines denote the $r^{-1}$ decay that is present for $ r < \xi_K$\footnote
{Note that these power-laws are not easily reproduced in a plot of the logarithmic derivative of the same data. Further work to pin down the precise nature of the correlation function is required.}.
We see that for $\lambda = 0.2$ the crossover from a $r^{-1}$ behaviour to a $r^{-2}$ decay is approximately shifted from $r \approx 7$ to $r\approx 16$
and the thermally induced exponential suppression of the spin spin correlation happens much later.
We can estimate the thermal cutoff scale by fitting exponentials (the yellow dashed lines), $e^{-\frac{r}{\xi_T}}$, to the tails of the plots for $\beta = 100$ and we find a consistent value of $\xi_T \approx 8.7$ for all values of $U$. 
A more detailed analysis of the temperature dependence for the point $U=2$ is found in \autoref{fig:exponential_Kondo}.
In \autoref{fig:Kondo_edge} (b) and (c) we can compare the temperature effects for $U=3$ and we see that at twice the temperature the exponential decay is not visible anymore.
Comparing the plots from (a) to (d) we can trace the shift of the cross-over from an $r^{-1}$ behaviour to an $r^{-2}$ decay with increasing $U$ and
therefore with the Kondo temperature.
\autoref{fig:Kondo_edge} (c) also shows the independence of the cross-over point of the algebraic decay due to the Kondo effect with respect to the external temperature.

\section{Summary}
We have studied a magnetic impurity coupled to a helical edge state as modeled by a Kane-Mele Hamiltonian on a slab geometry.
Due to time reversal symmetry the 
effective action of the impurity orbital (see \autoref{eq:dot_action} ) has precisely the same form as the generic SIAM such that the local physics is identical. In particular the 
Hubbard scale marks the appearance of a local moment which couples magnetically via the superexchange scale $J \propto V^2/U$ to the conduction electrons.
Below the Kondo temperature $T_K$, the magnetic moment is screened due to the formation of an entangled singlet state of the magnetic impurity and conduction electrons.
We have shown these commonalities numerically with the double occupancy,
the spin susceptibility and determined the Kondo temperature with a data collapse.

The differences to generic Kondo physics are non-local.
As shown in Ref. \cite{Maciejko2012}, if the interactions along the helical edge are not too strong, 
spin-flip single-particle backward scattering processes are expected to be irrelevant,
such that in the low temperature limit the conductance should reach the unitarity limit.
The mechanism which allows this to occur is the deflection of the edge state into the bulk, thus avoiding the Kondo singlet.
By computing the temperature dependence of the site 
resolved density of states by an analytic continuation of the impurity self-energy to real frequencies,
we were able to follow the temperature dependence of the edge mode and in particular its deflection into the bulk due to the emergence of the Kondo singlet. 
Complementary information on the extent of the Kondo singlet -- without resorting to an analytical continuation --
was also obtained  by computing the spatial dependence of the spin spin correlation functions between the local moment and conduction electrons.
At low temperatures the spin spin correlations do not extend significantly into the bulk and exhibit a power law decay along the edge.
In particular, as a function of temperature, we can observe the 
thermal cutoff scale $\xi_T \propto v_F \beta $ beyond which exponential decay sets in, as well as the characteristic cross-over 
scale around $\xi_K \propto v_F/T_K$ from an $r^{-1}$ to an $r^{-2}$ law.
This cross-over scale provides a measure of the Kondo screening cloud.
Of significant interest is an explicit calculation of the temperature dependence of the conductance along the edge.
In particular, in the local moment regime, where we observe no deflection of the edge state, one expects a decrease of the conductance due to back-scattering spin-flip processes off the impurity spin.
Below the Kondo scale, the deflection of the edge state along the new boundary of the system -- as defined by the topology of the Kondo cloud -- should restore the conductance to it's unitarity limit.

\section{Acknowledgments}
We especially thank M. Bercx and M. Weber for proof-reading the article.
We thank T. C. Lang for the figure of the honeycomb lattice.
We acknowledge support from DFG Grant No.~AS120/4-3.
We thank the LRZ Munich and the J\"{u}lich Supercomputing
Centre for generous allocation of CPU time.
DJL thanks the university of New Mexico --- where part of this work has been carried out --- for hospitality.
We thank Martin Hohenadler, Thore Posske and Masud Haque for fruitful discussions.

\appendix
\section{Analytical continuation}
\label{sec:analytical_continuation}
The analytic continuation of the Matsubara Green's function $G(n,k,\I \omega_n)$ to the real
frequency axis is a notoriously hard problem and requires potentially large amounts of computer
time. For Monte Carlo data, experience shows that the most reliable spectra can be obtained using
the stochastic maximum entropy method \cite{Sandvik98,Beach04} for analytic continuation. This
method, however, makes use of a Monte Carlo simulation to find the best spectrum and has to be
performed for every pair of indices $(n,k)$ individually. Clearly, this increases the computational
effort substantially.
  In order to circumvent this problem, we propose to adopt an idea that has been sucessfully used in
the context of the dynamical cluster approximation (DCA) \cite{Fuchs11}: Instead of the Green's function,
we calculate the analytic continuation of the self-energy.
The reason why this is beneficial is that the self-energy matrix has only one non-vanishing and
diagonal $2\times 2$ spin block $\Sigma(z)$ corresponding to the impurity site and leads to the special form of
Dyson's equation:
\begin{equation}
\label{eq:dyson_matrix_form}
\begin{split}
& \begin{pmatrix}
  G_{\vec{r},\vec{r}}(z) & G_{\vec{r},d}(z) \\
  G_{d,\vec{r}}(z) & G_{d,d}(z) \\
 \end{pmatrix}
=
 \begin{pmatrix}
  G^0_{\vec{r},\vec{r}}(z) & G^0_{\vec{r},d}(z) \\
  G^0_{d,\vec{r}}(z) & G^0_{d,d}(z) \\
 \end{pmatrix}
 +\\
 &\begin{pmatrix}
  G^0_{\vec{r},\vec{r}}(z) & G^0_{\vec{r},d}(z) \\
  G^0_{d,\vec{r}}(z) & G^0_{d,d}(z) \\
 \end{pmatrix} 
 \begin{pmatrix}
  0 & 0 \\
  0 & \Sigma(z) \\
 \end{pmatrix}
 \begin{pmatrix}
  G_{\vec{r},\vec{r}}(z) & G_{\vec{r},d}(z) \\
  G_{d,\vec{r}}(z) & G_{d,d}(z) \\
 \end{pmatrix}.
\end{split}
\end{equation}
Obviously, the knowledge of $\Sigma(z)$ on the real axis is therefore sufficient to obtain the full
interacting Green's function matrix $G(\omega+\I \eta)$ for the whole lattice as the noninteracting
Green's function $G^0$ can be calculated exactly with little difficulty by virtue of the resolvent
formalism.

Solving equation \ref{eq:dyson_matrix_form} for $\Sigma(z)$ yields 
\begin{equation}
\Sigma(z) = G^0_{d,d}(z)^{-1} - G_{d,d}(z)^{-1}.
\end{equation}
As $G_{d,d}(\I \omega_n)$ can be calculated in CT-INT, we can therefore obtain $\Sigma(\I
\omega_n)$. In order to analytically continue $\Sigma(z)$ to the real axis, we have to study its
asymptotic behaviour for large frequencies. Starting from the asymptotic series for
$G^{(0)}_{d,d}(\I \omega_n)$
\begin{equation}
\label{eq:g_asymptotics}
G^{(0)}_{d,d}(\I \omega_n) = \sum_{k=1}^{\infty} \frac{a^{(0)}_k}{(\I \omega_n)^k},
\end{equation}
we obtain through inversion:
\begin{equation}
\begin{split}
\Sigma(\I \omega_n) &= \left(a_2 - a_2^{0}\right)  + \frac{1}{\I \omega_n} \frac{ a_2^2 -
\left(a_2^{0}\right)^2 + a_1 a_3 -a_1 a_3^0 }{a_1}\\
 &+ \mathcal{O}\left(\frac{1}{(\I
\omega_n)^2}\right).
\end{split}
\end{equation}
This result can be obtained by truncating equation \ref{eq:g_asymptotics} at different orders and
one indeed finds out that higher terms of the Green's function's asymptotic series\footnote{Note that
the leading constant of the Green's function is generated by the canonical anticommutation relation
of $d^\dagger$ and $d$ and is therefore $a_1=a_1^{0}=1$.} do not contribute
to the first two terms of the self-energy.

In order to employ the stochastic maximum entropy method for $\Sigma(z)$ directly, we introduce a
slightly different quantity as already shown in reference \onlinecite{Fuchs11}:
\begin{equation}
\label{eq:self_energy_transformation}
\Sigma'(z) = \frac{ \left[ \Sigma(z) -  \left(a_2 - a_2^{0}\right) \right] a_1 }{ a_2^2 -
\left(a_2^{0}\right)^2 + a_1 a_3 -a_1 a_3^0 }.
\end{equation}
This quantity has exactly the same properties as the Green's function itself, namely that its
asymptotic series starts with $\frac{1}{\I \omega_n}$, the corresponding spectral function has a sum
rule $\int \D \omega A_\Sigma(\omega) = \pi$ and that it does not have a constant term.

In principle, these properties could be corrected for in the maximum entropy procedure but the
quantities $a_2$ and $a_3$ can only be obtained up to a statistical errorbar and therefore the
correct inclusion of these errors is very cumbersome. Performing the transformation
\ref{eq:self_energy_transformation} is therefore a very straightforward procedure, as the thoroughly
bootstrapped covariance matrix of $\Sigma'$ will contain all uncertainties stemming from the CT-INT
calculation.

The calculation of the constants in the asymptotic series of the self-energy is a straightforward
calculation of moments %
\footnote{
    For this, remember that the $\alpha$th moment of the spectral function can be obtained by the
expression
\begin{equation}
\intop \nolimits 
\mathrm{d}\omega
\omega^\alpha A_{d,d}(\omega) = (-1)^\alpha 
<
\left[ \left[ d^\dagger, H\right]_{-,\alpha} ,d \right]_+ 
>,
\end{equation}
with the recursive definition
\begin{equation}
\left[A,B\right].
\end{equation}
}%
 of the spectral function $A(\omega)$ corresponding to the Green's function
$G_{d,d}$ and yields for $A_{\Sigma'}^\sigma(\omega)$
\begin{equation}
a_2-a_2^0 = U \thavg{d_{-\sigma}^\dagger d_{-\sigma}} - \frac{U}{2}.
\end{equation}
\begin{equation}
\begin{split}
&\frac{ \left(a_2^0\right)^2 - a_2^2 + a_1 (a_3 - a_3^0)}{a_1} =\\ &UV \left(
\thavg{a_{\vec{0},-\sigma}^\dagger d_{-\sigma} } - \thavg{d_{-\sigma}^\dagger a_{\vec{0},-\sigma}}
\right) \\ &+ U^2 \thavg{ d_{-\sigma}^\dagger d_{-\sigma} } - U^2 \thavg{ d_{-\sigma}^\dagger d_{-\sigma} }^2.
\end{split}
\end{equation}
At half filling, $\thavg{d^\dagger_\sigma d_\sigma}=\frac{1}{2}$, so the constant term $a_2-a_2^0$
of the self-energy vanishes.

\section{Diagonal $G_{d,d}$ due to TRI}
\label{sec:Gisdiagonal}
Since $[H,T] = 0$, the operators $H$ and $T$ have a common basis in which they are diagonal.
In the case that $T^2 = -\mathbb{1}$ holds (which is the case if the total spin is of the half-integer type)
we have for every state $\vec{e_1} = \ket{a}$ an orthogonal state $\vec{e_2} = T \ket{a}$.
In these basis states $T$ has the matrix representation
\begin{equation}
 T = \begin{pmatrix}
      0 & \mathbb{1} \\
      -\mathbb{1} & 0
     \end{pmatrix}
\end{equation}
where $\mathbb{1}$ denotes the identity matrix acting in the respective sub-sector of the state. By that notation we have essentially just renamed
the states of the Hilbert space.
$T$ is diagonalized by 
\begin{equation}
 U  = \frac{1}{\sqrt{2}}\begin{pmatrix}
       \mathbb{1} & i\mathbb{1} \\
       i\mathbb{1} & \mathbb{1} 
      \end{pmatrix}
\end{equation}
with two eigenvectors named $\ket{+}$ and $\ket{-}$.
$U$ also block-diagonalizes $H$ and hence the Green's function matrix is block-diagonal in the $\ket{+}$ and $\ket{-}$ basis
since propagation with $H$ will not generate matrix elements between the othogonal sub-spaces.
\begin{equation}
 G = \begin{pmatrix}
      G^{++} & 0 \\
      0 & G^{--} 
     \end{pmatrix}.
\end{equation}
Transforming $G$ back to the original basis we have
\begin{equation}
\begin{split}
 G &= U G U^\dagger = \frac{1}{2}
 \begin{pmatrix}
  G^{++} + G^{--} & i G^{++} -i G^{--} \\
  -i G^{++} + i G^{--} & G^{++} + G^{--}
 \end{pmatrix}.
 \end{split}
 \label{eq:finalG}
\end{equation}
To show that the off-diagonals of this matrix vanish in the sub-space of this matrix where the impurity lives,
we consider the impurity Green's function $G_{d,d}$ in the $\ket{+}$ and $\ket{-}$ basis.
Then, using time-reversal symmetry,
\begin{equation}
 T H T^{-1} = H
\end{equation}
we have for $\tau > 0$,
\begin{equation}
 \begin{split}
  G_{d,d}^{s,s}(\tau) & = \text{Tr}\left( \E^{-\beta H} d^\dagger_s(\tau) d_{s} \right) \\
  &= \text{Tr}\left( \E^{-\beta T H T^{-1}} d^\dagger_s(\tau) d_{s} \right) \\
  &= \text{Tr}\left( T \E^{-\beta H} T^{-1} \E^{-\tau H} d^\dagger_s \E^{\tau H} d_{s} \right) \\
  &= \text{Tr}\left(\E^{-\beta H} \E^{-\tau H} T^{-1}d^\dagger_s T \E^{\tau H} T^{-1} d_{s} T \right)
 \end{split}
\end{equation}
using $T^{-1} d_s T = s d_{-s}$ (the impurity lacks a momentum quantum number) we have
\begin{equation}
 \begin{split}
  G_{d,d}^{s,s} & = s^2 \text{Tr}\left(\E^{-\beta H} d^\dagger_{-s}(\tau) d_{-s} \right) \\
  & = G_{d,d}^{-s,-s}.
  \end{split}
\end{equation}
Therefore we have $G^{+,+}_{d,d} = G^{-,-}_{d,d}$ thereby ensuring that the off-diagonals in \autoref{eq:finalG} vanish.
This in turn gives the diagonality of an impurity Green's function $G_{d,d}$ just due to time reversal symmetry.
\bibliography{references}
\end{document}